%% file: Fermi_mn.tex
\documentclass[useAMS,usegraphicx,usenatbib]{mn2e}
\newcommand{\mybibstyle}{mymn}

\usepackage{color}
\newcommand{\rev}[1]{{#1}}
\usepackage{aas_macros}
\usepackage{deluxetable}

\usepackage[dvips, bookmarks, pdfborder={0 0 0.1}]{hyperref}

\usepackage{grffile}

\usepackage{amssymb,amsmath}
\usepackage[section] {placeins}
\usepackage{multirow}

\newcommand{\pow}[1]{\times 10^{#1}}
\newcommand{\msun}{\mathrm{M}_\odot}
\newcommand{\bbbar}{$\mathit{b\bar{b}}\;$}
\newcommand{\bbbarnoc}{$\mathit{b\bar{b}}$}
\newcommand{\mumu}{$\mathit{\mu^+\mu^-}\;$}
\newcommand{\tautau}{$\mathit{\tau^+\tau^-}\;$}
\def\lsim{ \lower .75ex \hbox{$\sim$} \llap{\raise .27ex \hbox{$<$}} }
\def\gsim{ \lower .75ex \hbox{$\sim$} \llap{\raise .27ex \hbox{$>$}} }
\graphicspath{{show/}}

\begin{document}

\title[Extended Gamma-ray Emission in Clusters]{Constraining Extended Gamma-ray Emission from Galaxy Clusters}

\author[J. Han et al.]
{Jiaxin Han,$^{1,2,3}$\thanks{jxhan@shao.ac.cn} Carlos S. Frenk,$^{3}$ 
Vincent R. Eke,$^{3}$ Liang Gao,$^{4,3}$
\newauthor Simon D. M. White,$^{5}$ Alexey Boyarsky,$^{6,7}$ 
Denys Malyshev$^7$ and Oleg Ruchayskiy$^8$\\
$^1$Key Laboratory for Research in Galaxies and Cosmology,
Shanghai Astronomical Observatory, Shanghai 200030, China\\
$^2$Graduate School of the Chinese Academy of Sciences,
19A, Yuquan Road, Beijing 100049, China\\
$^3$Institute of Computational Cosmology, Department of Physics,
University of Durham, Science Laboratories,\\ South Road, Durham DH1
3LE\\
$^4$Partner Group of the Max Planck Institute for Astrophysics,
National Astronomical Observatories, \\Chinese Academy of Sciences,
Beijing, 100012, China\\
$^5$Max-Planck Institute for Astrophysics,
 Karl-Schwarzschild Str. 1, D-85748, Garching, Germany\\
$^6$ Instituut-Lorentz for Theoretical Physics, Universiteit Leiden,
Niels Bohrweg 2, Leiden, The Netherlands\\
$^7$ Bogolyubov Institute of Theoretical Physics, Kyiv, Ukraine\\
$^8$ CERN Physics Department, Theory Division, CH-1211 Geneva 23,
Switzerland}

\maketitle
 
\begin{abstract}
\input{abstract_\version}

\end{abstract}


\input{main_\version}

\end{document}

%% file: abstract_v3f.tex
Cold dark matter models predict the existence of a large number of
substructures within dark matter halos.  If the cold dark matter
consists of weakly interacting massive particles, their annihilation
within these substructures could lead to diffuse GeV emission that
would dominate over the annihilation signal of the host halo.  In this
work we search for GeV emission from three nearby galaxy clusters:
Coma, Virgo and Fornax. We first remove known extragalactic and
galactic diffuse gamma-ray backgrounds and point sources from the
Fermi 2-year catalog and find a significant residual diffuse emission
in all three clusters. We then investigate whether this emission is
due to (i) unresolved point sources; (ii) dark matter annihilation; or
(iii) cosmic rays (CR).  Using 45 months of Fermi-LAT data we detect
several new point sources (not present in the Fermi 2-year point
source catalogue) which contaminate the signal previously analyzed by
Han et al. Including these and accounting for the effects of
undetected point sources, we find no significant detection of extended
emission from the three clusters studied. Instead, we determine upper
limits on emission due to dark matter annihilation and cosmic rays.
For Fornax and Virgo the limits on CR emission are consistent with
theoretical models, but for Coma the upper limit is a factor of 2
below the theoretical expectation. Allowing for systematic
uncertainties associated with the treatment of CR, the upper limits on
the cross section for dark matter annihilation from our clusters are
more stringent than those from analyses of dwarf galaxies in the Milky
Way. Adopting a boost factor of $\sim 10^3$ from subhalos on cluster luminosity 
as suggested by recent theoretical models, we rule out the thermal cross section for supersymmetric dark matter particles for masses as large as 100 GeV (depending on the annihilation channel).

%% file: main_v3f.tex
\section{Introduction}

The existence of dark matter (DM) in the universe has so far only been
deduced from its gravitational effect, due to the lack of
electromagnetic interactions of the DM with itself or with baryonic
matter. There are several elementary particle candidates for DM in
various extensions of the standard model of particle
physics \citep{Bertone_04}. Weakly interacting massive particles or WIMPs (with a
self-interaction cross-section at roughly the weak scale) are one class of the
popular dark matter candidates. 
These particles could be related to the electroweak symmetry breaking
which is currently being explored by experiments at the LHC. For
example, within the framework of the minimal supersymmetric standard
model (MSSM), the lightest neutralino emerges as a candidate WIMP that
is stable over cosmological timescales and can annihilate into
standard model particles.
WIMPs behave as  
cold dark matter since their primordial velocity
dispersion is negligible. High resolution N-body simulations 
show that cold dark matter halos contain a population of self-bound
substructures (subhalos) whose number decreases with increasing
subhalo mass as $N\propto M^{-\alpha}$ with $\alpha \approx 1.9$
\citep{Diemand07,Volker,Gao}

Much effort has been devoted to the search
for 
WIMPs either directly or indirectly. Direct detection involves
identifying the rare events of DM scattering off ordinary matter or
searching for new particles near the weak scale at the LHC. Indirect
detection involves looking for the annihilation or decay products of
dark matter in cosmic rays and gamma rays. In particular, pair
annihilation produces gamma-ray photons at a rate proportional to the
square of the dark matter density, which then propagate, almost
without absorption, to the observer. In this case, the Galactic centre
should be the brightest gamma-ray source on the sky \citep[][and
references therein]{Volker}. Extended emission (distinct from the
central point source) was reported from the central $1^\circ$ around
the Galactic centre by \citet{GalCen1,GalCen}.\footnote{See also a preliminary result by the Fermi-LAT
  collaboration \citep{FermiGalCen}.}
 This emission has been interpreted as a signal from dark matter
annihilation. \rev{There is, however, a strong ambiguity in
modeling this region of the Galaxy. Its angular size is comparable with
the PSF size of the Fermi LAT at these energies, and
the galactic diffuse background there is known to be complicated and highly
non-uniform. In particular, it was demonstrated in \citet{GalCen3}
that although an extra diffuse component improves the quality of fit, the
radial profile of the ``extended emission" is fully consistent with that of
known Fermi point sources and therefore the emission could all originate from
point sources at the Galactic centre\citep[see also][]{Abazajian12}.} 
An intriguing aspect of a DM explanation for the gamma-ray emission
from the Galactic centre is that the inferred particle mass of around
10 GeV is also the mass claimed to be required to explain other data,
such as the synchrotron emission from the Milky Way's radio filament
\citep{DM_radio} and the ``WMAP Haze'' \citep{Haze1,Haze2,Haze3}, as
well as signals from the direct detection experiments DAMA/LIBRA
\citep{DAMA}, CoGeNT \citep{CoGeNT1,CoGeNT2} and CRESST-II
\citep{CRESST}.  {These signals, however, could be in tension with other
direct detection experiments, such as CDMS~\citep{CDMS} and
XENON-100~\citep{Xenon100}, although optimistic arguments also exist (e.g.,\citet{Collar1,Collar2}. }
We refer the reader to \citet{DanRev} for review.

It has recently been reported that the $\gamma$-ray emission from the
region around the Galactic centre exhibits a line-like excess at
energies $\sim 130$~GeV
\citep{Brigmann:2012,weniger12,tempel12,Finkbeiner12}. The
intepretation of this signal as arising from dark matter particles,
however, is controversial \citep[see][]{boyarsky12}.

 
Targeting the entire sky rather than the Galactic centre in searching for
annihilation radiation may seem a good strategy since this takes advantage of
the large-scale distribution of dark matter while avoiding some of the
uncertainties arising from the astrophysical modelling of galactic gamma-ray
sources. However, the fact that we are located near the centre of the Galactic
halo and that most of the annihilation emission outside the Galactic centre is
produced by dark matter substructures \citep{Diemand07,Volker} results in a
gamma-ray map from annihilation that is almost uniform on large scales. This
makes detection within the Milky Way halo a difficult task, exacerbated by the
additional uncertainty of having to model the extragalactic background, which
is more important on large scales~\citep{FermiLarge,robust}.

Dwarf galaxies are the most DM-dominated objects known, are relatively 
free from astrophysical contamination and appear compact on the 
sky. They are therefore promising targets to search for DM annihilation 
radiation. Recent joint analyses of eight to ten dwarf galaxies 
\citep{Dwarf,DwarfFermi} resulted in no significant detection but have 
began to rule out the canonical annihilation cross-section of $3\times 
10^{-26}\rm{cm}^3\rm{s}^{-1}$ for DM masses below $\sim30-40$~GeV.

Galaxy clusters are the most massive virialized DM structures in the 
universe and are also good targets for indirect DM searches. The 
presence of a large population of DM substructures (or subhalos) 
predicted by numerical simulations further enhances the detectability of 
DM in clusters. Although the total mass within subhalos amounts to 
only 10 to 20 percent of the total halo mass, the density enhancement 
within subhalos can boost the total cluster annihilation luminosity by 
a factor as high as 1000 when extrapolated down to a subhalo mass 
limit of one Earth mass, the fiducial cutoff in the primoridal power 
spectrum of density fluctuations for a typical 100 GeV 
WIMP 
\citep{Gao,Pinzke2}. As the distribution of subhalos is much less 
concentrated than that of the smooth main halo, the total annihilation 
emission from clusters is predicted to be extended. Thus, attempts to 
detect DM annihilation assuming a point source or NFW-squared 
profile could miss most of the signal. In fact, just such a search 
using the 11-month Fermi-LAT data has yielded no significnat detection 
of emission from six clusters \citep{FermiCluster}.

Using the 45-month data, we consider possible contributions from 
cosmic ray (CR) induced gamma-ray emission and from DM 
annihilation. For the former (which can be as high as, or higher than 
the emission from cluster DM annihilation 
\citep{CRvsDM,Pinzke1,Pinzke2}, we adopt the semi-analytic method 
developed by \citet{Pinzke1}. For the later, we adopt the model 
proposed by \citet{Gao} for the cluster DM annihilation profile. We 
provide constraints on both the CR and DM components for the three 
galaxy clusters analyzed by Han et al. (2011): Coma, which is 
predicted to have the highest signal-to-noise according to 
\citet{Gao}, and Fornax and Virgo which are predicted to have the 
lowest astrophysical contamination according to \cite{Pinzke2}. 


The current paper replaces an earlier version by a subset of the 
authors \citep[][arXiv:1201.1003]{Han_v1}. After submission of that version, it was 
pointed out to us that a number of point sources are present in the 
full three-year LAT data which were not detected significantly in the 
data used for the ''official'' Fermi point source catalogue available at 
the time of our analysis, the LAT 2-year point source catalogue (2FGL; 
\citet{2FGL}). We now carry out our own point source detection in the 
regions of interest and find several new point sources.\footnote{{We notice that several new point sources in Virgo are also identified in a concurrent paper\citep{NewPT} and are found to reduce
the significance of DM-like emission in the cluster, consistent with what we find here.}
} We account for these new detections in our analysis, as well as for the fact that a 
significant part of the ``smooth" extragalactic background is 
contributed by point sources below the detection threshold; this 
alters the noise properties of this background. Both changes reduce 
the significance of the diffuse components apparently detected in the 
first version of our paper, so that we can now reliably only place 
upper limits. 

{\cite{Huang} have recently reported a failure to detect significant DM 
annihilation emission from a combined analysis of eight galaxy 
clusters. Our work differs from theirs in several respects: firstly, 
we assume a DM annihilation profile based on high resolution 
cosmological simulations \citep{Gao}; secondly, we assess the impact 
of cosmic rays in the detection of dark matter; and finally, we 
include in our sample the Virgo cluster which turns out to be the best 
candidate. The constraints we set on the annihilation cross-section 
are consistent with those of \citet{Huang}.}

The paper is organized as follows. In section~\ref{sec_data_analysis}
we describe the data and provide an overview of the models of the
Virgo, Fornax and Coma galaxy clusters regions used in the analysis
(see Table~\ref{table_property}). The specification of the
non-standard components of the models (dark matter and cosmic rays
brightness profiles) is provided in Sec.~\ref{sec_model}. The
constrains on CR emission and DM annihilation that we obtain are
summarized in section~\ref{sec_results} and discussed in
Sec.~\ref{sec_discussion}.

The cosmological parameters used in this work are the same as those 
assumed by \citet{Gao}: 
$\Omega_m=0.25$, $\Omega_\Lambda=0.75$, $h=0.73$.

\label{sec_data_analysis} 
\subsection{Data preparation}\label{sec_prep} 
We analyze the first 45 months of data (04/08/2008 to 20/05/2012) from 
the Fermi-LAT, 
\footnote{http://fermi.gsfc.nasa.gov/cgi-bin/ssc/LAT/LATDataQuery.cgi} 
trimmed with the cuts listed below, to select high quality photon 
events. This typically results in $\sim 10^5$ photons within a radius 
of 10 degrees around each cluster, while the expected number of 
annihilation photons is of the order of $10^2$ according to 
Fig.~\ref{fig_profile}. The most recent instrument response function 
(IRF), P7CLEAN\_V6, is adopted for the analysis, in accordance with 
our event class selection.\footnote{We also tried using P7SOURCE\_V6 
IRF and Event Class 2 data. The results are consistent with those 
presented in this paper.} The resulting gamma-ray images for the three 
clusters are shown in the left panel of Fig.~\ref{f_components} for 
Virgo and in Fig.~\ref{fig_CMaps} for Coma and Fornax.

\begin{tabbing} 
\hspace{5.5cm}\=\kill 
Minimum Energy     \>    100 MeV\\ 
Maximum Energy     \>   100 GeV\\ 
Maximum zenith angle\footnotemark  \> 100 degrees\\ 
Event Class\footnotemark  \> 3 (P7CLEAN) \\ 
DATA-QUAL\footnotemark  \> 1 \\ 
LAT CONFIG\footnotemark  \> 1 \\ 
ABS (ROCK ANGLE)\footnotemark \> $<52$ degrees\\ 
ROI-based zenith angle cut \> yes \\ 
\end{tabbing}

\footnotetext[5]{ZENITH ANGLE (degrees): angle between the 
reconstructed event direction and the zenith line (originates at the 
centre of the Earth and passes through the centre of mass of the 
spacecraft, pointing outward). The Earth's limb lies at a zenith angle 
of 113 degrees.} 
\footnotetext[6]{EVENT CLASS: flag indicating the probability of the 
event being a photon and the quality of the event reconstruction.} 
\footnotetext[7]{DATA-QUAL: flag indicating the quality of the LAT 
data, where 1 = OK, 2 = waiting review, 3 = good with bad parts, 0 = 
bad} 
\footnotetext[8]{LAT-CONFIG: flag for the configuration of the lat (1 
= nominal science configuration, 0 = not recommended for analysis)} 
\footnotetext[9]{ROCK ANGLE: angle of the spacecraft $z$-axis from the 
zenith (positive values indicate a rock toward the north).}

We list the basic properties of the three clusters in Table~\ref{table_property}. 
\begin{table*} 
\caption{Basic Properties of Target Clusters}\label{table_property} 
\begin{tabular}{|c|c|c|c|c|} 
\hline   & Coma & Fornax & Virgo (M87) \\ 
\hline RA (deg) & 194.9468 & 54.6686 & 187.6958  \\ 
\hline DEC (deg) & 27.9388 & -35.3103 & 12.3369 \\ 
\hline $D_A$ (Mpc)\tablenotemark{a} & 95.8 & 17.5 & 16.8 \\ 
\hline $M_{200}$ ($\msun$)\tablenotemark{b} & 1.3e15 & 2.4e14 & 7.5e14 \\ 
\hline $r_{200}$ (deg)\tablenotemark{b} & 1.3 & 4.1 & 6.2 \\ 
\hline $\mathcal{J}_{NFW}$\tablenotemark{c} & 5.9e-5 & 4.1e-4 & 1.2e-3 \\ 
\hline Enhancement due to subhalos within $r_{200}$ \tablenotemark{d} & 1.3e3 & 6.5e2 & 1.0e3\\ 
\hline 
\end{tabular} \\ 
\tablenotetext{a}{Angular diameter distance, from the NASA 
extragalactic database for Coma and Fornax, and from \citet{TS84} for Virgo.}
\tablenotetext{b}{Cluster halo mass defined as the mass within the radius, 
$r_{200}$, within which the average density equals 200 times the 
critical density of the universe. Values for Coma and Fornax are taken from 
\citet{Pinzke2}, while the value for Virgo is taken from \citet{TS84}.} 
\tablenotetext{c}{Integrated coefficient, 
$\mathcal{J}_{int}=\int_{\Delta \Omega} J d\Omega$, over the solid 
angle spanned by the cluster virial radius, assuming a smooth NFW 
density profile.} 
\tablenotetext{d}{Enhancement to the total annihilation 
luminosity within the virial radius due to substructures, 
extrapolated to a subhalo mass limit of $10^{-6}\msun$. 
{Note this factor scales with the minimum subhalo mass as $M_{cut}^{-0.226}$ \citep{Volker}.}} 
\end{table*} 

\subsection{Maximum-likelihood fitting}\label{sec_ML}

We use the \texttt{pyLikelihood} tool shipped with the Fermi Science 
Tools software package (version v9r27p1-fssc-20120410) to perform a 
maximum likelihood (ML) analysis \citep{EGRET}. After applying 
appropriate data cuts, as described in section~\ref{sec_prep}, we bin 
the data into 0.1 degree-wide pixels 
and 30 logarithmic energy bins within a radius of 10 degrees around 
each cluster. This large radius is chosen to account for the large LAT 
PSF size at low energies ($4\sim10$ degrees at 100 MeV\footnote{The LAT PSF size scales 
roughly as $E^{-0.8}$, so at 1~GeV it is $\sim 1$deg}). An exposure cube 
is computed around each cluster covering 25 degrees in radius and the 
30 energy bins, using the \texttt{gtexpcube2} tool.

In the standard Fermi likelihood analysis, the photon counts within each pixel are treated assuming Poisson 
statistics for each energy bin to calculate the likelihood. The 
best-fit parameters are obtained when the likelihood for the entire data set 
is maximized. The significance of a given component of interest 
(e.g. DM or CR) from the ML fitting is quantified by the likelihood 
ratio statistic,

\begin{equation}\label{eq_TS} 
TS=-2\ln(L_0/L), 
\end{equation} 
where $L$ is the maximum likelihood for the full model and $L_0$ is 
the maximum likelihood for the null hypothesis, i.e, the model {\em 
without} the component of interest. According to Wilk's theorem, this test 
statistic, $TS$, approximately follows a $\chi^2$ distribution when the null 
hypothesis is true, with one degree of freedom for our case where the normalization is the only 
extra parameter in the alternative model. The probability that a given value of $TS$ arises purely 
from fluctuations of the null hypothesis is: 
\begin{equation}\label{eq_P_TS} 
P=\int_{TS}^{\infty}\frac{1}{2}\chi_1^2(\xi)d\xi 
=\int_{\sqrt{TS}}^{\infty}\frac{e^{-x^2/2}}{\sqrt{2\pi}}dx. 
\end{equation} 
The factor $\displaystyle\frac{1}{2}$ comes from the constraint that 
the normalization parameter be non-negative. The significance of a 
detection can thus be quoted as $\sqrt{TS}\sigma$ (one-sided Gaussian 
confidence). Upper limits on the extra normalization parameter $N$ are 
obtained by searching for a null hypothesis $L_0'$ where $N$ in the 
full model is constrained to be equal to the upper limit, $N_{UL}$, so 
that ${\rm ln}(L_0'/L)=-1.35$, corresponding to the 95\% confidence 
interval. 

\subsection{Model} 
For the analysis we constructed a model to fit the data including all known foreground 
and background emission, as well as DM and CR components, as 
appropriate. We include all the point sources from 2FGL within a 
radius of 15 degrees from the cluster centre in the model, plus the 
most recent galactic (GAL) and extragalactic (EG) diffuse emission 
given by the template files \texttt{gal\_2yearp7v6\_v0.fits} and 
\texttt{iso\_p7v6clean.txt}. Additionally, we have searched the 
45-month data for new point sources; we detect several of them within 
the cluster region (see Appendix~\ref{sec_NewPT} for more detail) and 
these are also included in our model. The normalization of the GAL and 
EG diffuse components are allowed to vary during the fitting. Within 
the cluster virial radius there are two 2FGL point sources and one 
newly detected point source in Fornax, six 2FGL, including the central 
AGN \citep[M87;][]{M87Fermi}, plus four newly detected ones in Virgo.  We 
allow the normalization and power-law spectral index of these thirteen 
point sources to vary freely. In addition, the parameters of all 
sources with variability index greater than 50 located within 10 
degrees of the cluster centres are allowed to vary. Parameters for the 
other point sources are fixed as in the 2FGL catalog. From now on we 
refer to the model with GAL, EG and the known point sources as the 
``base model''.

A DM annihilation surface brightness template (given by the dimensionless factor J, see Eqn.~\ref{eq_J} in Sec.~\ref{sec_DM_annihilation_template}) 
is generated for each 
cluster out to a 15 degree radius by summing up both the contribution 
from a smooth NFW profile and the contribution from subhalos. This J-map is 
used to fit for extended cluster annihilation emission. For the point 
source model, the integrated factor $\mathcal{J}_{NFW}$ (see  
Eqn.~\ref{eq_JNFW}) is used to derive an annihilation cross-section 
from the fitted total flux. 
Similarly, a CR photon template is generated for each 
cluster out to three times the cluster virial radius, where the 
surface brightness has dropped to below $10^{-5}$ of the central value 
and beyond which the model is not reliable. Images for various model 
components are shown in Fig.~\ref{f_components} taking Virgo as an 
example. We discuss these templates in more details in Sec.~\ref{sec_model}.

In the traditional Fermi analysis, the EG template is treated as a 
smooth component where all emission below the nominal point source 
detection limit is assumed to come from a smoothly distributed diffuse 
component. In this work, we also consider a more realistic one where a 
fraction is assumed to be contributed by fainter point sources with a 
number-flux relation which extrapolates smoothly from that measured 
for brighter sources. In this case the photon counts within a given 
pixel are no longer Poisson-distributed since the photons arrive in 
packets.  In principle, one can use the full distribution of photon 
counts from a population of randomly placed point sources to calculate 
the likelihoods $L$ and $L_0$, but Eqns.~\ref{eq_TS} and~\ref{eq_P_TS}, and 
the corresponding discussion, are not affected.  However, since the 
full distribution of photon counts in this case (Han et.al., in prep.) is 
complicated and difficult to implement in the likelihood analysis, 
instead of recalculating $L$ and $L_0$, in this work we use Monte-Carlo 
simulations to re-evaluate the distribution of $TS$ for the more 
realistic background model and provide corrections to the results of 
the standard analysis where needed.

\section{Modeling gamma-ray emission in clusters }\label{sec_model}

\begin{figure*} 
\includegraphics[scale=0.3]{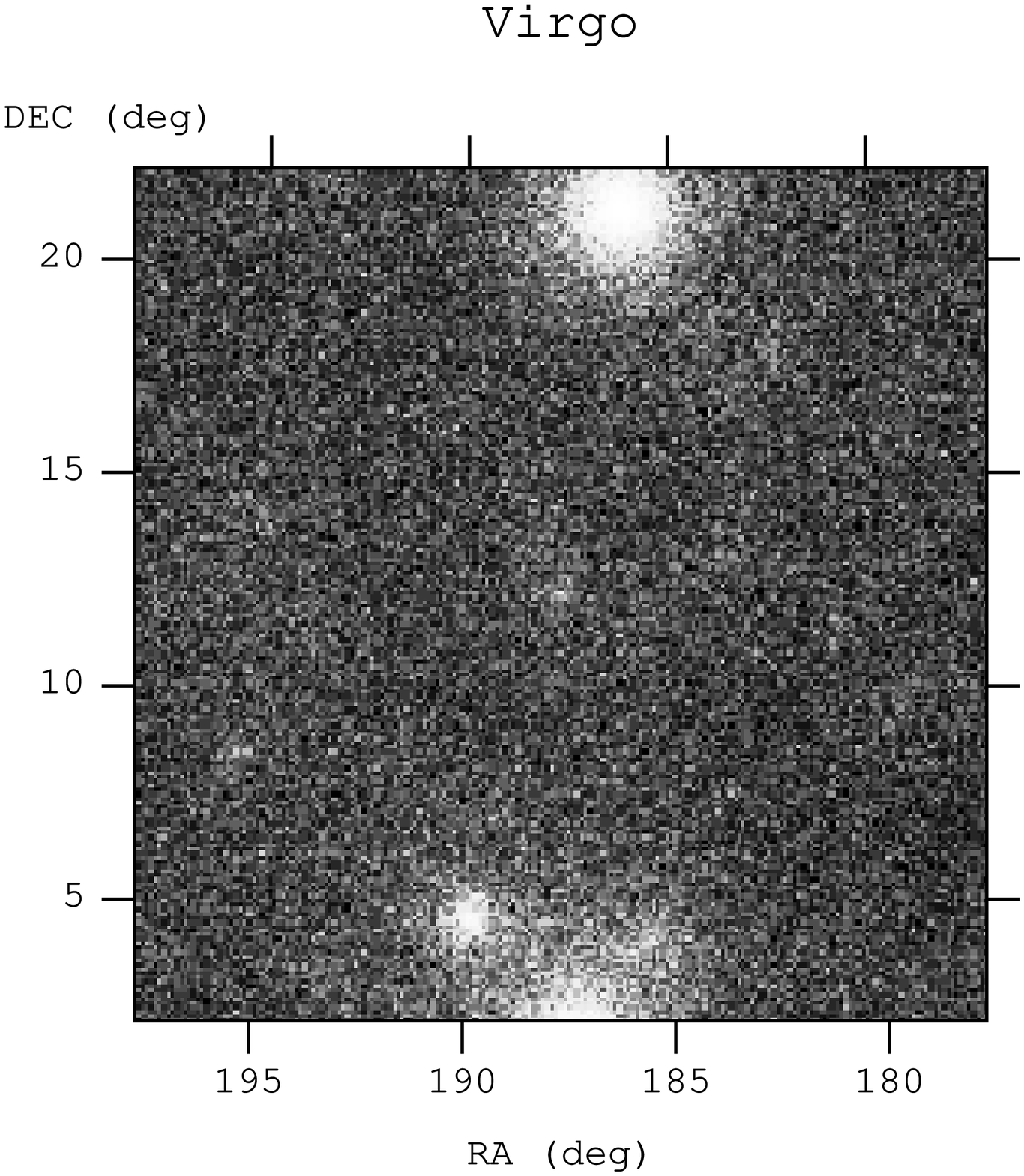}%
\includegraphics[scale=0.5]{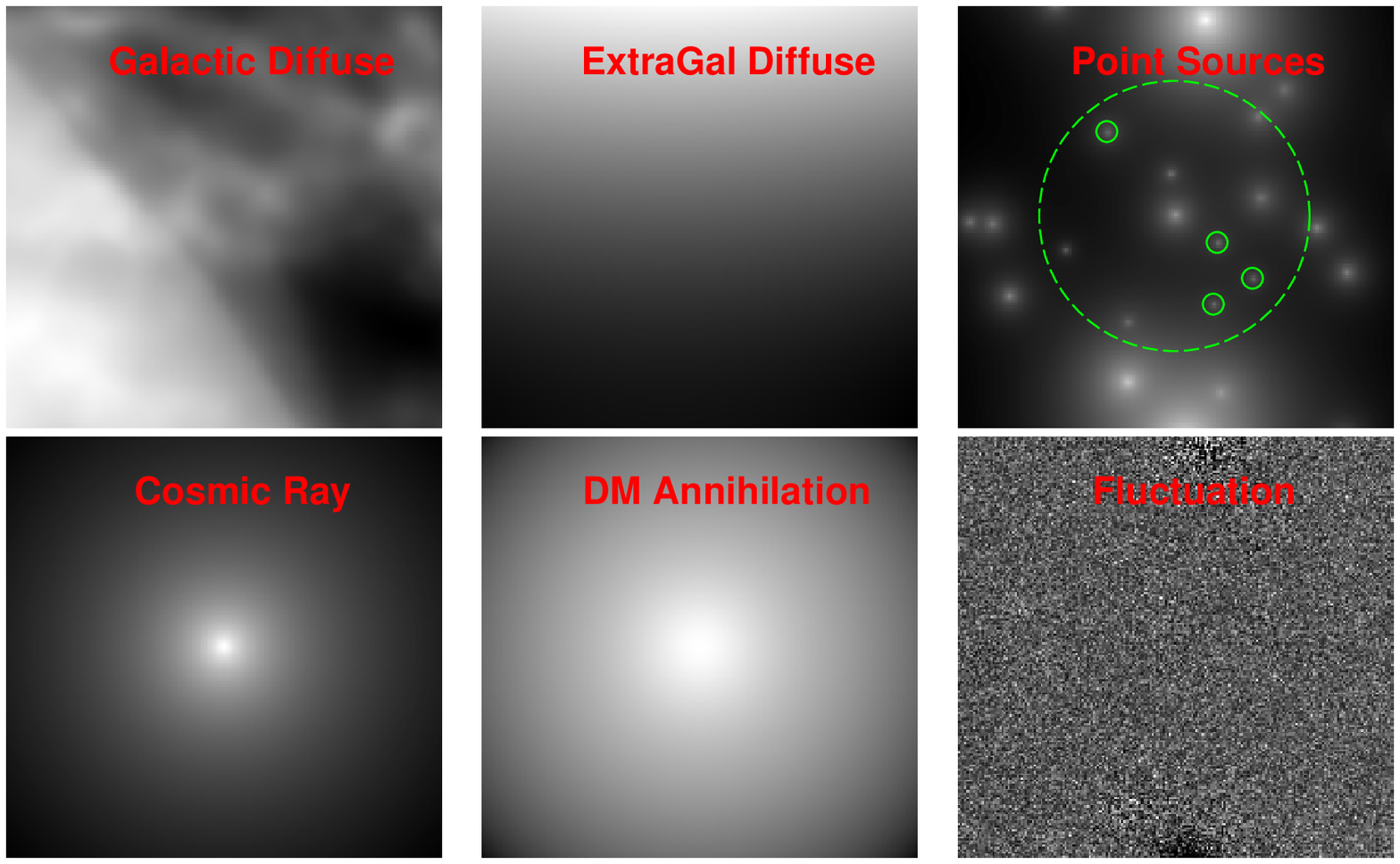} 
\caption{Decomposition of the Fermi-LAT image in the region of the 
Virgo cluster into model components. The observed photon count image 
from 100~MeV to 100~GeV is shown on the left. The right panels show the 
integrated image over the same energy range for the various model 
components: galactic diffuse emission, extragalactic diffuse emission, 
detected point sources, cosmic-ray photons and DM annihilation emission, 
as labeled. The green dashed circle in the ``Point Sources'' panel marks the 
virial radius of the cluster. The small circles mark the newly detected point sources 
which are not present in the 2FGL. The ``Fluctuation" panel shows the 
residual image for our best-fit DM model. The images have been 
enhanced individually in colour space for contrast. Note the apparent 
structure in the extragalactic component which is due to different 
exposure times at different positions.} 
\label{f_components} 
\end{figure*}

We model the observed gamma-ray emission in clusters with several 
components as shown in Fig.~\ref{f_components}: the galactic 
foreground (GAL), the extragalactic background (EG), emission from 
known point sources, DM annihilation and CR-induced emission. The GAL 
and EG diffuse emission are given by the most recent templates, 
\texttt{gal\_2yearp7v6\_v0.fits} and \texttt{iso\_p7v6clean.txt}, 
which can be obtained from the Fermi-LAT data server, while the point 
sources include those from the LAT 2-year point source catalogue, 2FGL 
\citep{2FGL}, as well as several, newly detected by us, in the 
45-month data. In addition, an improved EG model which includes a population of 
un-detected sources is also analyzed. We now describe in detail our models for DM 
annihilation and CR emission.

\subsection{Dark matter annihilation emission} 
\label{sec_DM_annihilation_template} 
The gamma-ray intensity along the line-of-sight due to DM annihilation 
is given by:

\begin{equation}\label{eq_I} 
I=\frac{1}{8\pi}\sum_f\frac{dN_f}{dE}<\sigma_f v>\int_{l.o.s.} 
(\frac{\rho_\chi}{M_\chi})^2(l)dl, 
\end{equation} 
where $M_\chi$ is the DM particle mass; $\rho_\chi$ the density of DM; 
 $\displaystyle\frac{dN_f}{dE}$ 
the particle model dependent term giving the differential number of 
photons produced from each annihilation event as a function of energy, 
$E$, in a particular annihilation channel, $f$; and $<\sigma_f v>$ is the 
velocity-averaged cross-section (or annihilation rate) for that channel, 
which is predicted to be constant in the low velocity limit appropriate to 
present-day cold DM particles (see e.g., \citet{SUSYDM}). The 
line-of-sight integration of the 
density squared is often expressed in terms of a dimensionless factor, 
\begin{equation}\label{eq_J} 
J=\frac{1}{8.5\rm{kpc}}(\frac{1}{0.3\rm{GeV}/\rm{cm}^3})^2\int_{l.o.s.} 
\rho_\chi^2(l)dl. 
\end{equation} 
If the source size is much smaller than the instrumental beam size, a 
point source approximation is applicable. In this case, the 
integration of $J$ over a large enough solid angle, $\Delta\Omega$, is used to 
determine the total flux for the point source, 
$\mathcal{J}_{int}=\int_{\Delta \Omega} J d\Omega$.

The cluster annihilation emission is modeled with the extended profile 
suggested by \citet{Gao}. However, for part of the analysis and for 
comparison purposes, we will also use the point source approximation 
which, although inappropriate, has been employed in all previous 
analysis of Fermi-LAT data from clusters.  We shall refer to models 
that assume these two profiles respectively as EXT and PT. If the 
cluster follows a smooth NFW profile, then its integrated $J$ factor 
which determines the total annihilation flux can be found as 
\begin{equation}\label{eq_JNFW} 
\mathcal{J}_{NFW}=\frac{4\pi}{3} \rho_s^2 r_s^3\frac{1}{D_A^2} \times 
\frac{1}{8.5\rm{kpc}}(\frac{1}{0.3\rm{GeV}/\rm{cm}^3})^2. 
\end{equation} 
Here $D_A$ is the angular diameter distance to the cluster and 
$\rho_s$ and $r_s$ are the characteristic density and radius of the NFW 
profile. They are related to halo concentration, $c$, and virial radius 
through the relations, 
$\rho_s=\dfrac{200}{3}\dfrac{c^3\rho_c}{{\rm ln}(1+c)-c/(1+c)}$ and 
$r_s=r_{200}/c$, with $\rho_c$ the critical density of the universe, 
$r_{200}$ the cluster virial radius within which the average density 
is $200\rho_c$ and the concentration parameter, $c$, is given by the 
following mass-concentration relation: 
\begin{equation} 
c=5.74(\frac{M_{200}}{2\times 10^{12}h^{-1}\msun})^{-0.097} 
\end{equation} 
\citep{M-c}. Here, $M_{200}$ is the mass enclosed within $r_{200}$. 
Extrapolating to a cutoff mass of $10^{-6}\msun$, the existence of 
subhalos will increase this flux by a factor 
\begin{equation} 
b(M_{200})=\mathcal{J}_{sub}/\mathcal{J}_{NFW}=1.6\times 
10^{-3}(M_{200}/\msun)^{0.39} 
\end{equation} 
\cite{Gao}. Using the results of the simulations by these authors, 
the surface brightness profile of subhalo emission can be 
fitted within $r_{200}$ by the following formula: 
\begin{equation} 
J_{sub}(r)=\frac{16b(M_{200})\mathcal{J}_{NFW}}{\pi 
\ln(17)}\frac{D_A^2}{r_{200}^2+16r^2}\;\;\;\;\; (r \leq r_{200}). 
\end{equation} 
Below we fit the subhalo emission surface brightness beyond the 
virial radius and extrapolate to several times the virial radius using 
an exponential decay, 
\begin{equation} 
J_{sub}(r)=J_{sub}(r_{200})e^{-2.377(r/r_{200}-1)}\;\;\;\;\; (r \geq 
r_{200}). 
\end{equation} 
The total annihilation profile is the sum of the contributions from a 
smooth NFW profile and the subhalo emission. This is completely 
dominated by subhalo emission except in the very centre of the 
cluster. We show the total annihilation profile and its decomposition 
into main halo and subhalo contributions in the left panel of 
Fig.~\ref{fig_profile}, taking Virgo as an example. This profile is 
further inflated after convolution with the LAT point spread function.

We consider three representative annihilation channels, namely into $b-\bar{b}$, 
$\mu^+-\mu^-$ and $\tau^+-\tau^-$ final states. The annihilation spectrum 
is calculated using the DarkSUSY package \citep{DarkSUSY}, 
\footnote{http://www.darksusy.org.} which 
tabulates simulation results from 
PYTHIA.\footnote{http://home.thep.lu.se/~torbjorn/Pythia.html} We also 
include the contribution from inverse Compton (IC) scattered photons 
by energetic electron-positron pairs produced during the annihilation 
process, following the procedure described in \citet{Pinzke2}. In 
general, three external energy sources are involved in the dissipation 
and scattering of the injected electrons from annihilation: the Cosmic 
Microwave Background (CMB), infrared to UV light from stars and dust, 
and the interstellar magnetic field. However, as shown by 
\citet{Pinzke2}, the latter two components are expected to be 
important only in the inner region of clusters ($<0.03r_{200}$), 
corresponding to less than 0.2 degrees for our three 
clusters. Including them would introduce a position-dependent 
component to the annihilation spectrum, so for simplicity we only 
consider the contribution of CMB photons in the IC calculation. For 
the \bbbar channel, IC photons only contribute significantly to the 
low energy spectrum for relatively high neutralino mass, while for the 
leptonic channels, which have plenty of energetic electrons, the IC emission 
can completely dominate the annihilation emission over the full 
energy range of interest for the highest neutralino masses 
considered.

\begin{figure*} 
\includegraphics[width=\textwidth]{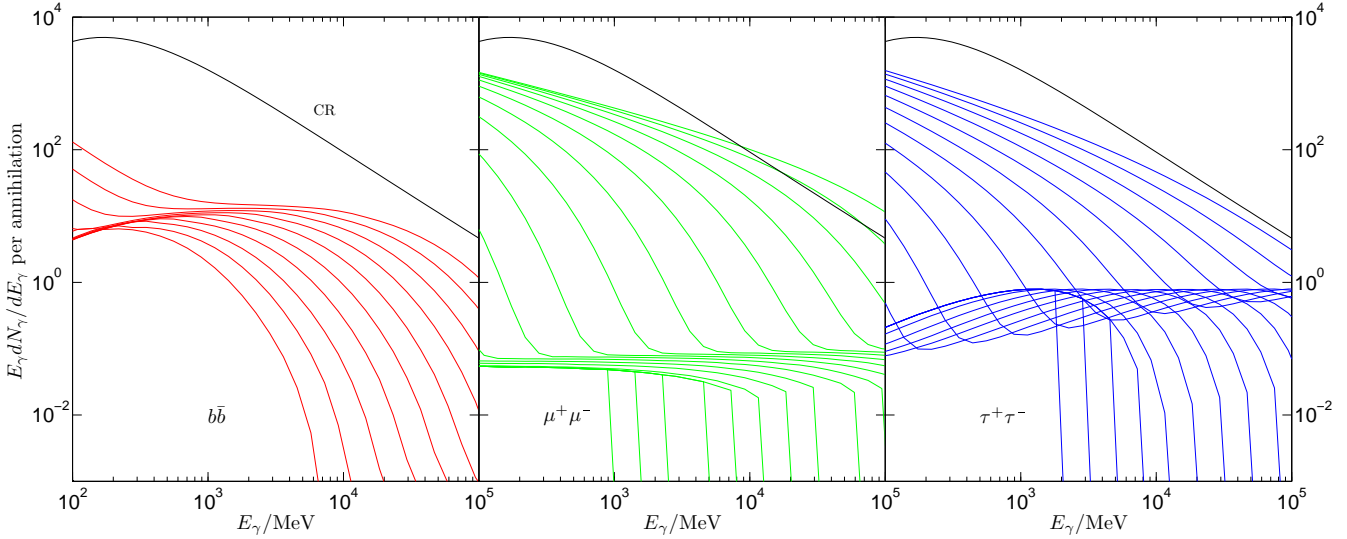} 
\caption{Photon yields for \bbbar (left), \mumu (middle) 
and \tautau (right) channels. 
Plotted are the total photon yields including continuum 
secondary photons, final state radiation from charged final state 
particles, as well as inverse Compton scattering of CMB photons by 
electron/positron pairs, for the mass range $10-1000$~GeV\ for the 
\bbbar channel, $1\rm{GeV}-10\rm{TeV}$ for the \mumu channel and 
$2\rm{GeV}-10\rm{TeV}$ for the \tautau channel. 
 The masses are sampled uniformly in a log scale. Note that each spectrum 
cuts off at an energy corresponding to the particle mass. For 
comparison, the black line in each panel shows the photon spectrum from cosmic ray 
induced photons with arbitrary normalization.}\label{fig_yield} 
\end{figure*}

We note that the electroweak corrections recently proposed by 
\citet{PPPC_theo} (see also \citet{PPPC_fit}) can bring visible 
differences to the leptonic channel spectra at high WIMP masses before IC 
scattering. However, since IC photons dominate at the high mass end 
and the electroweak correction only significantly changes the positron 
yields at low energy, thus having little effect on the IC spectrum, 
the electroweak correction to the total spectrum is still 
negligible. The total photon yields are shown in 
Fig.~\ref{fig_yield}. The almost flat spectrum with a cutoff around 
the energy corresponding to the WIMP mass comes from prompt 
annihilation emission including continuum secondary photons and final 
state radiation from charged final state particles. The low energy rise 
originates from IC scattered CMB photons.

\subsection{Cosmic-ray induced gamma-ray emission within clusters}

The cosmic ray induced gamma-ray emission is calculated following 
a semi-analytic prescription, derived from high resolution numerical 
simulations of galaxy clusters, that models cosmic ray physics self 
consistently \citep{Pinzke1}. The gamma-ray photon production rate (or source function) 
from pion decay is found to be separable into a 
spatial and a spectral part:

\begin{equation} 
q_{CR}(r,E)\equiv\frac{dN_{\gamma}}{dtdVdE}=A(r)s(E), 
\end{equation} 
where the spatial part, $A(r)$, is proportional to the square of the 
gas density profile multiplied by a slowly varying radial function 
parametrized by cluster mass. The spectral part, $s(E)$, is almost 
independent of cluster mass and has a power-law form, 
$dN_{\gamma}/d\ln(E_{\gamma})\propto E_{\gamma}^{-1.3}$, for the energy 
range $1\sim100$~GeV but flattens at low energies, as shown in 
Fig.~\ref{fig_yield}. We summarize the detailed form of $A(r)$ and 
$s(E)$ plus the gas density profile for the three clusters derived 
from X-ray observations in the Appendix.

\begin{figure*} 
\includegraphics[width=0.5\textwidth]{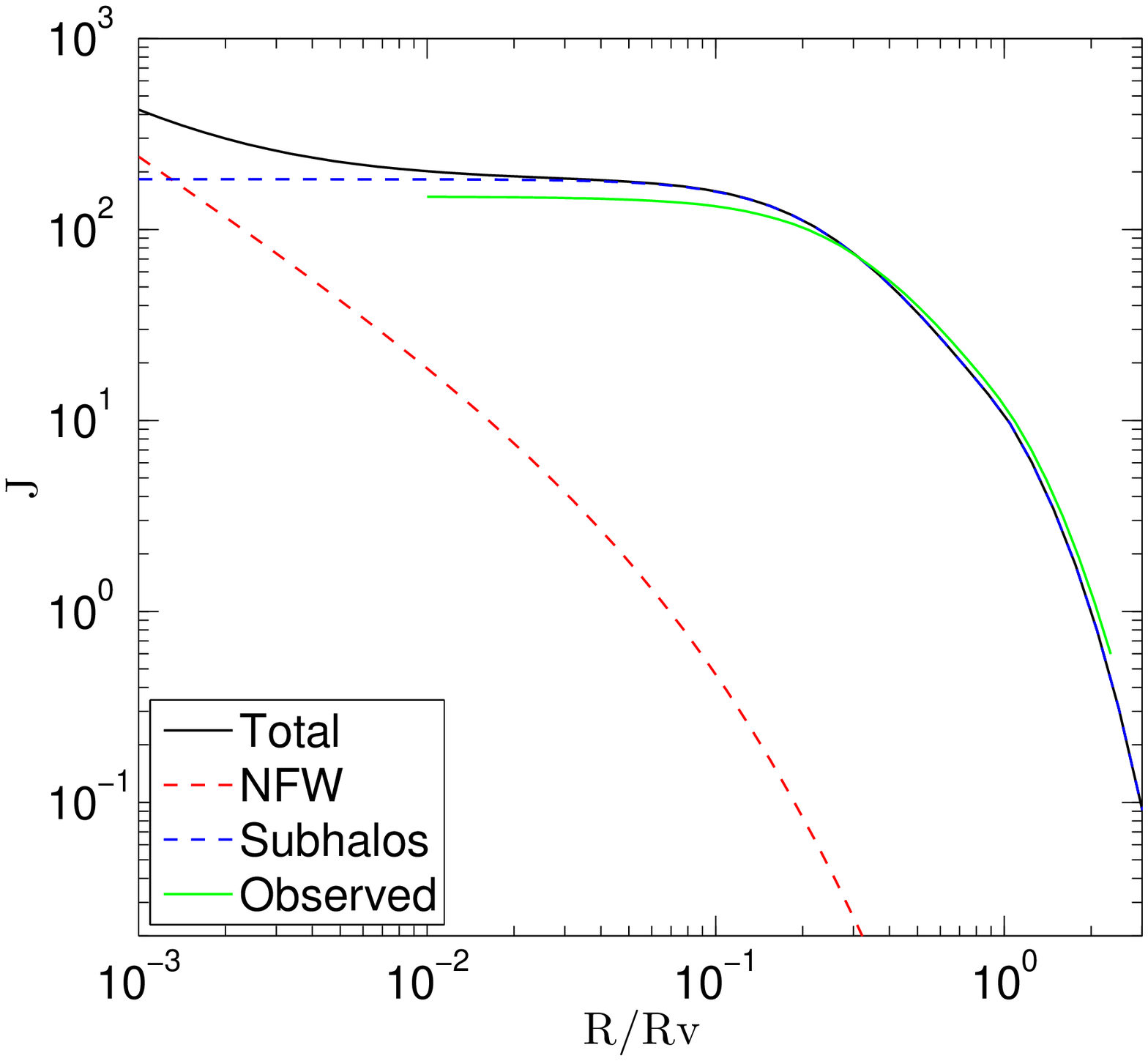}%
\includegraphics[width=0.5\textwidth]{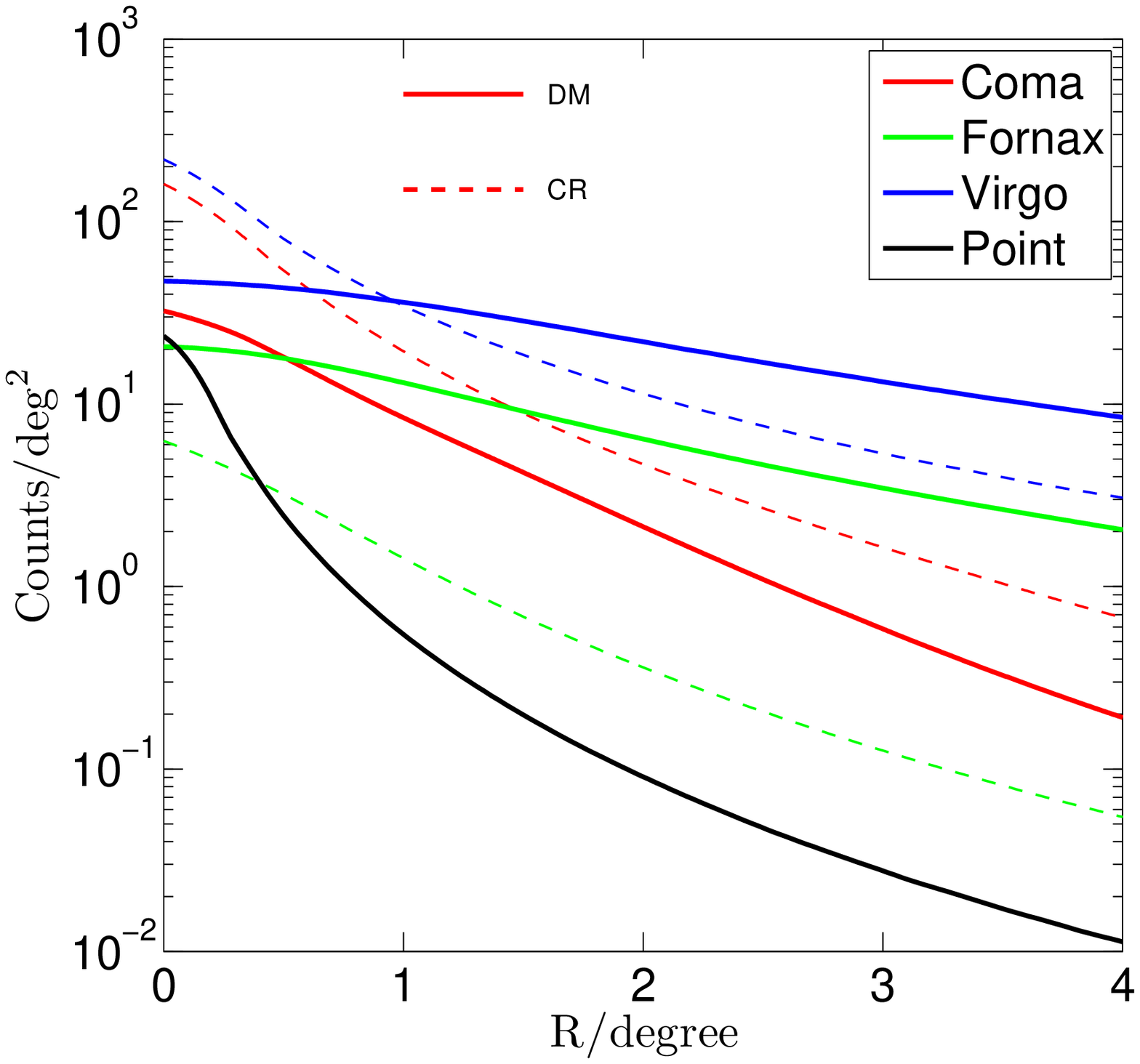} 
\caption{Cluster photon profiles. Left: theoretical and PSF-convolved 
J profile for Virgo. The total annihilation profile is shown as a 
black solid line and is decomposed into the smooth main halo part (red 
dashed line) and the subhalo part (blue dashed line). The green solid 
line shows the annihilation profile after PSF convolution, plotted 
down to an inner radius comparable to the pixel size of 
0.1~deg. Right: PSF-convolved photon profiles from annihilation 
(solid) and cosmic rays (dashed) for three clusters (indicated by 
different colours). Solid lines show the expected photon count 
profile for the extended DM annihilation model. Dashed lines show the 
expected cosmic-ray induced photon counts for the fiducial CR 
model. For comparison, we also plot the PSF-convolved profile for a 
central point source (black solid line) with arbitrary 
normalization. In both panels, a dark matter model with particle mass, 
$M\approx 100{\rm GeV}$, and annihilation cross-section, $<\sigma 
v>=3\times10^{-26} \rm{cm^3 s^{-1}}$, through the \bbbar channel is 
assumed. The PSF convolutions are done with the \texttt{gtmodel} 
tool in the Fermi Science Tools software package.}\label{fig_profile} 
\end{figure*}

The differential gamma-ray flux from this source function, 
$I_{CR}(r,E)$, is simply the integral of $q_{CR}(r,E)$ along the 
line-of-sight.  This prescription is derived from the average emission 
profile for a sample of simulated clusters for a realistic choice of 
parameter values (e.g., for the maximum shock acceleration efficiency, 
$\zeta_{p,max}$). In addition to the uncertainties in the model 
parameters there is also uncertainty in the observationally derived 
halo mass and gas density profile. In this work, we simply assume that 
the shape of $q_{CR}(r,E)$ is given by the model described above and 
account for the uncertainty in the model parameters, as well as sample 
variance with an additional normalization parameter, $\alpha_{CR}$, so 
that, 
\begin{equation} 
I_{CR}(r,E)=\alpha_{CR} \int_{l.o.s} \frac{q_{CR}(r,E)}{4\pi} dl. 
\end{equation} 
We take $\alpha_{CR}=1$ as our fiducial CR model and also consider the 
case when $\alpha_{CR}$ is fitted from the actual gamma-ray data as an 
optimal model. In the right panel of Fig.~\ref{fig_profile} we 
compare the CR profile for the fiducial model to the expected DM 
annihilation profile within our three clusters, assuming a fiducial DM 
particle model with particle mass, $M\approx 100{\rm GeV}$, annihilating 
through the \bbbar channel with cross-section, $<\sigma 
v>=3\times10^{-26} \rm{cm^3 s^{-1}}$. In general the CR emission is 
more centrally concentrated than the annihilation profile since the CR 
trace the gas profile. It can be readily seen that Fornax has a 
particularly low CR level while Coma is CR dominated. Coma has steeper 
profiles due to its larger distance and hence smaller angular size.

\section{Results}\label{sec_results} 
\subsection{Constraints on CR emission}

With all the model components defined above, we first proceed with ML 
fitting for a model with no DM annihilation but with cosmic rays, the 
``CR-only'' model hereafter. Note that the GAL and EG backgrounds, as 
well as the nearby point sources, are always included in the 
analysis, as described in section~\ref{sec_ML}. The results for the 
CR-only model fits are listed in Table~\ref{table_CR}. The fitted 
CR levels all agree within a factor of three with the theoretical 
predictions. While Fornax is most consistent with no CR emission due to its 
intrinsically low CR level, the derived upper limit for Coma already 
rules out the fiducial value at 95\% confidence. 
\begin{table*} 
\caption{Fits to the CR-only Model}\label{table_CR} 
\begin{tabular}{|c|c|c|c|c|c|c|} 
\hline  & $\alpha_{CR,fit}$\tablenotemark{a} & $\alpha_{CR,UL}$\tablenotemark{b} & $ F_{CR,UL}\tablenotemark{c}$~(ph $\cdot$ cm$^{-2}$ s$^{-1}$) & $TS$ & $TS_{corrected}$\tablenotemark{d} & $\alpha_{CR,UL,corrected}$\tablenotemark{e}\\ 
\hline Coma & $0.3\pm 0.1$ & 0.5 & 2.4e-09 & 5.2 & 2.6 & 0.6 \\ 
\hline Fornax & $0.9\pm 2$ & 4.8 & 1.8e-09 & 0.2 & 0.1 & 6.4\\ 
\hline Virgo & $0.6\pm 0.3$ & 1.2 & 2.1e-08 & 8.4 & 2.8 & 1.6 \\ 
\hline 
\end{tabular}\\ 
\tablenotetext{a}{Best fit normalization ($\alpha_{CR,fit}=1$ is 
the theoretical prediction)} 
\tablenotetext{b}{95\% upper limit (UL) on the normalization} 
\tablenotetext{c}{95\% upper limit on the CR induced gamma-ray flux 
from 100~MeV to 100~{\rm GeV}} 
\tablenotetext{d}{TS after allowing for undetected point sources; see Section~\ref{sec_correct} for details} 
\tablenotetext{e}{Upper limit on the normalization factor  after 
  allowing for undetected point sources; see Section~\ref{sec_correct} for 
  details} 
\end{table*} 
\subsection{Constraints on DM annihilation}

Given the low significance of the CR detection in the CR-only model, it is not 
safe simply to adopt the best fit $\alpha_{CR}$ values for further 
extraction of the DM signal. Instead, we consider the following 
four families of cosmic ray models in the presence of a DM component:

\begin{description} 
\item[Fiducial-CR model.] The CR level is fixed to the theoretical 
expectation, $\alpha_{CR}=1$. Since this value exceeds our derived upper 
limit for Coma, we exclude Coma from further discussion of this family. 
\item[Optimal-CR model.] The  CR level is taken as  the best-fit value listed in 
Table~\ref{table_CR}. 
\item[Free-CR model.] The normalization of the CR level is left as 
a free parameter in the fit. 
\item[No-CR model.] No CR emission is considered, only DM. 
\end{description} 
For each family, both point source (PT) and extended (EXT) profiles 
are considered for the DM component (the former merely for comparison 
with earlier work). Note that when calculating the $TS$ for DM, the 
null hypothesis refers to the full model excluding only the DM 
component, or equivalently, to the base model plus a CR component 
modelled according to one of our four families of CR models. We 
show results for the \bbbarnoc,  \mumu and \tautau  DM annihilation channels.

For none of the combinations of DM and CR models considered here, do 
we obtain a detection of DM at high significance in any of the three 
clusters. The highest significance is obtained for Virgo for the 
\bbbar channel in a DM model that has a particle mass of 30~GeV and 
the EXT profile, in the absence of CR. In this case, we find 
$TS=11.6$, corresponding to 3.4$\sigma$.  This reduces to $2.6\sigma$ 
in the Free-CR model and to less than $1\sigma$ in the Fiducial-CR 
model.

\begin{figure} 
\includegraphics[width=0.5\textwidth]{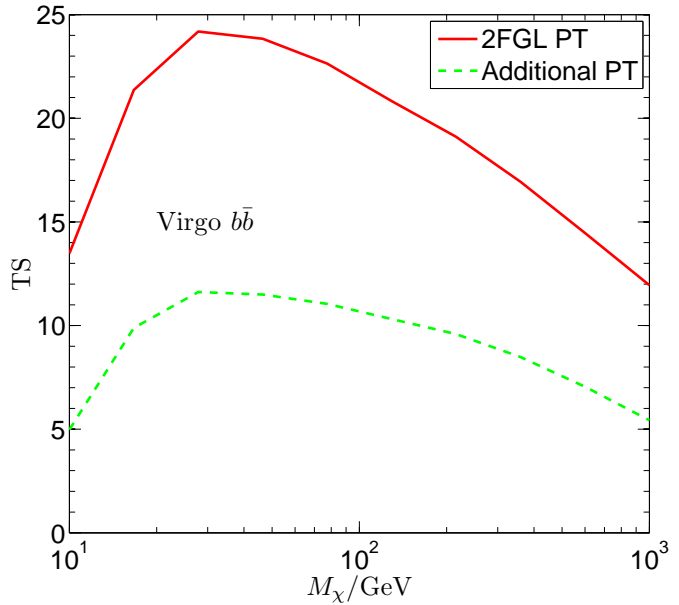} 
\caption{The significance of a DM component in Virgo, with \bbbar 
  final states, in the absence of CR. The solid line shows the TS when 
  only 2FGL sources are included in the model, while the dashed line 
  shows the case when the four new point sources that we have detected 
  are also  included.}\label{fig_TSCmp} 
\end{figure}

The value of $TS=11.6$ for the no-CR model for Virgo can be compared 
with the value of $TS=24$ reported in an earlier version of this paper 
(arXiv:1201.1003v.1) from a similar analysis of the 2-year Virgo data 
(see Fig.~\ref{fig_TSCmp}). The decrease in significance is entirely 
due to the subtraction of the new point sources which we have detected 
in 
the Virgo region and which were not catalogued in the 2FGL. These 
previously undetected sources happen to lie within the virial radius 
of Virgo and can mimic the extended emission expected from DM 
annihilation.  In fact, fits assuming an EXT profile but a power-law, 
rather than a DM annihilation spectrum, result in a similarly high 
significance detection, $TS=21$, and a best-fit spectral index 
$\Gamma=-1.9$. This is the typical spectral index of Fermi point 
sources (including the newly detected ones).  The preference for a 
30~GeV DM particle mass in the DM fits reflects a preference for a 
$\Gamma=-1.9$ spectrum around 1~GeV, the energy scale from which most 
of the significance arises.

In fact, the significance of the Virgo detection is further reduced 
when we take account in the analysis of a possible undetected point 
source population, as we shall do in Appendix~\ref{sec_CPD}. Thus, in 
what follows we use our analysis exclusively to set upper limits on 
the flux and annihilation cross-section.

\subsubsection{The \bbbar channel}\label{sec_bb} 
\begin{figure*} 
\includegraphics[width=0.5\textwidth]{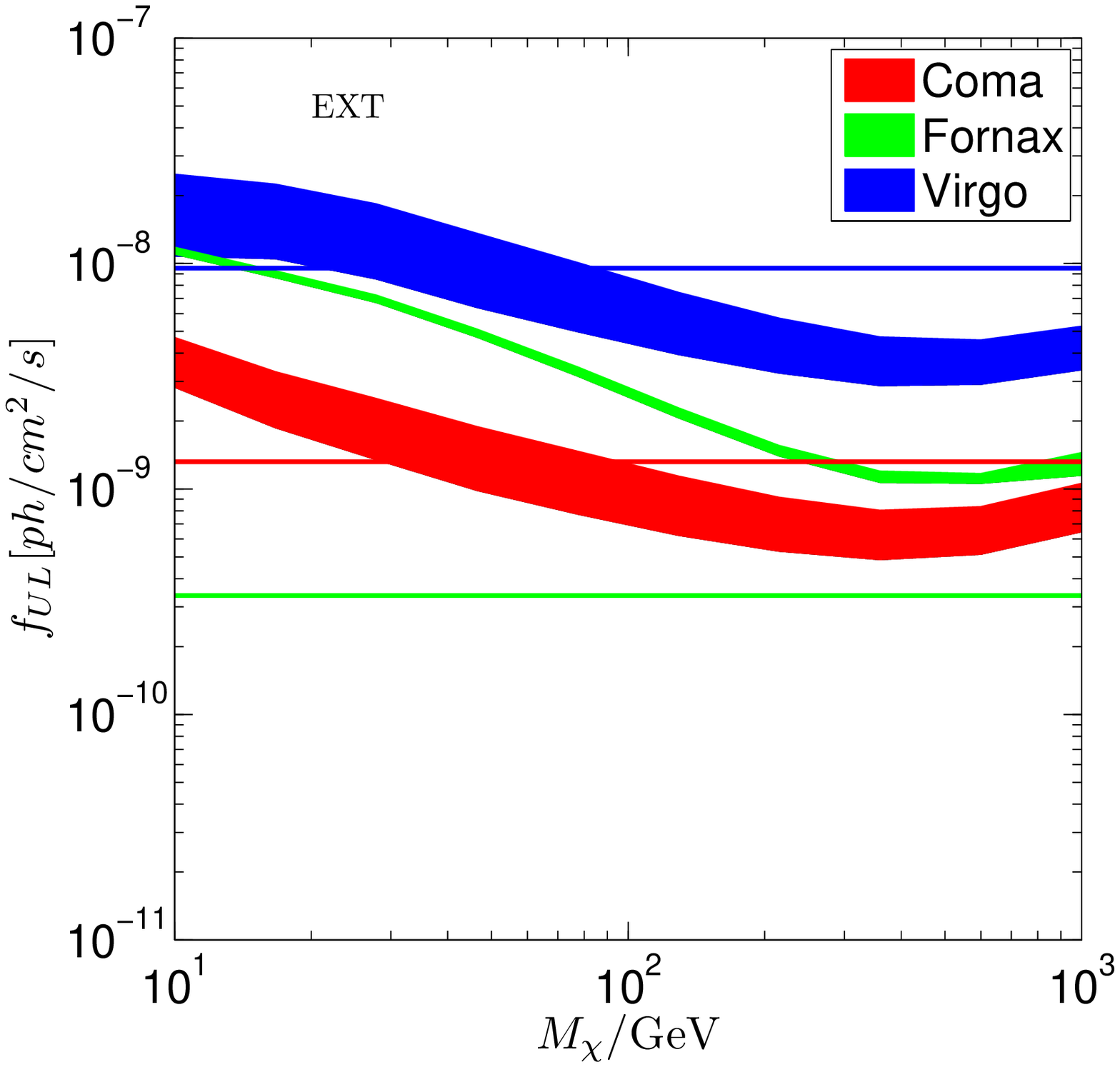}%
\includegraphics[width=0.5\textwidth]{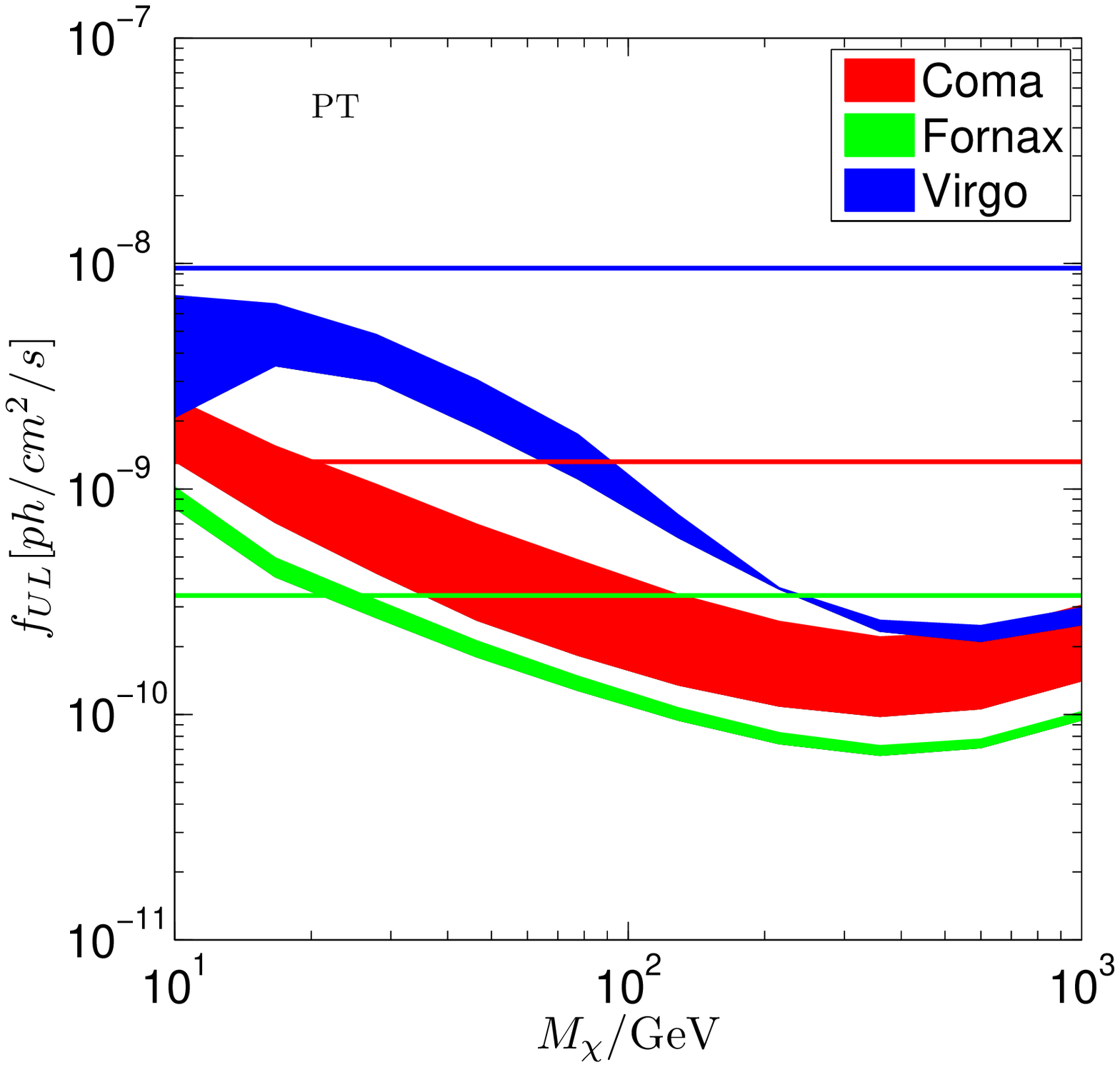} 
\caption{DM annihilation flux upper limits for the \bbbar channel. The 
stripes are defined by the minimum and maximum upper limits given by 
the four CR model families, with different colours corresponding to 
different clusters, as indicated in the legend. Left and right panels 
are the results for the EXT and PT profiles respectively. For each 
cluster, a solid line of the corresponding colour shows the optimal CR 
flux. }\label{fig_Fluxbb} 
\end{figure*}

In Fig.~\ref{fig_Fluxbb} we show the 95\% confidence upper limits on 
the DM annihilation flux and compare them to the CR levels. For each 
cluster, the coloured stripes are defined by the minimum and maximum 
upper limits corresponding to the four families of CR models. The optimal 
CR levels in the three clusters are all comparable to the fitted DM 
flux, and the DM flux upper limits for the four different CR models 
vary only within a factor of two, with the No-CR and Fiducial-CR 
\footnote{In Coma, where the Fiducial-CR model is ruled out, 
the Optimal-CR model yields the lowest upper limit.} cases 
predicting the highest and lowest upper limits. The left and right 
panels show the results for the EXT and PT models respectively; the PT 
models always have lower flux upper limits than the extended models.

\begin{figure} 
\begin{minipage}{0.5\textwidth} 
\includegraphics[width=\textwidth]{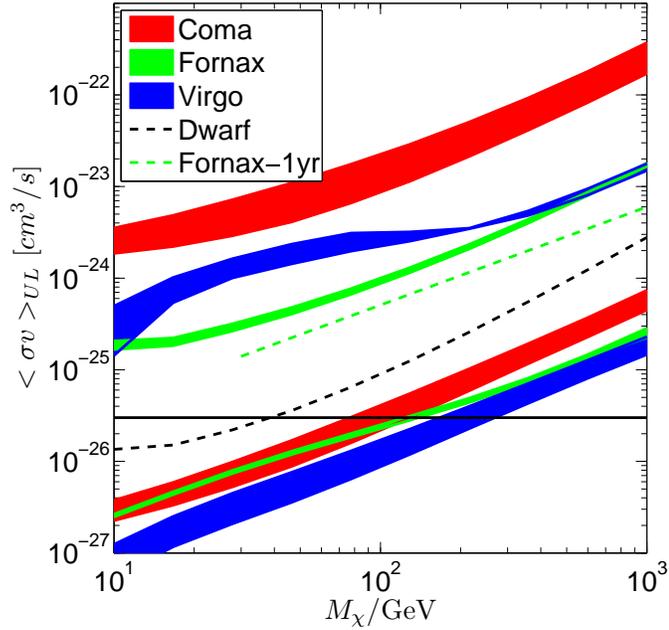} 
\caption[]{Upper limits for the DM annihilation cross-section in the 
\bbbar channel. The different colours represent the three clusters, with 
the stripes spanning the range between the minimum and maximum upper 
limits given by the four different ways of treating the CR 
component. The three highest stripes show the PT model constraints and 
the three lowest the EXT model constraints. We also plot with dashed 
lines constraints\footnote{The ``Fermi-1yr" constraint is 
only reproduced schematically, by reading out several data points from 
the original plot in the reference.} from a joint analysis of the 
Milky Way dwarf galaxies \citep[][black dashed line]{Dwarf} and 
previous constraints from the 11-month Fermi-LAT data for Fornax 
\citep[][green dashed line]{FermiCluster} assuming these authors' optimistic 
value for the total enhancement  due to subhalos, which gives the tightest 
constraint. The black solid line indicates the canonical thermal 
cross-section of $3\times 10^{-26} \rm{cm^3 s^{-1}}$. }\label{fig_SigVbb} 
\end{minipage} 
\end{figure}

The flux upper limits are translated into cross-section upper limits 
in Fig.~\ref{fig_SigVbb}, using Eqn.~\ref{eq_I}. These are also 
shown as coloured regions reflecting the variation in the different 
treatments of CR. Although the predicted flux upper limits decrease 
slowly with DM particle mass and remain within the same order of 
magnitude for the mass range considered, the resulting cross-section 
upper limits increase by a factor of 100 from low to high particle 
mass.  This is because low mass particles correspond to higher DM 
number densities (the $\rho_\chi^2/M_\chi^2$ factor in 
Eqn.~\ref{eq_I}) for a given mass density, so to obtain the same 
flux level, the required cross-section must be smaller for low mass 
particles. With an enhancement of order $10^3$  due to subhalos, a much 
lower cross-section is needed (by a factor of at least 100) for 
extended annihilation models to achieve a slightly higher flux upper 
limit than point source models.

Our cross-section limits drop below the fiducial thermal cross-section 
of $3\times 10^{-26}{\rm cm}^3\rm{s}^{-1}$ for $M_\chi\lsim 
100$GeV. Of the three clusters, Virgo has the highest flux upper 
limits but it still places the tightest constraints on the 
annihilation cross-section.  Our limits are much lower than those in 
the 11-month Fermi-LAT analysis by \citet{FermiCluster}, where the 
tightest constraint came from Fornax for a much lower assumed subhalo 
contribution of $\sim 10$. Our limits are also tighter than that from 
a joint analysis of the dwarf satellites of the Milky Way by 
\citet{Dwarf}. \footnote{If systematic uncertainties in the halo mass 
  parameters assumed by \citet{Dwarf} are considered, the lower bounds 
  of their derived limits become comparable to our limits.} Note that 
the difference between our results and that of \citet{FermiCluster} 
comes mostly from different assumptions about the effect of subhalos, 
and only secondarily from the larger amount of data we have 
analysed. Also, note that in the analysis of \citet{Dwarf}, no boost 
from subhalos within the halo of dwarf galaxies was assumed.

\subsubsection{The \mumu channel} 
\begin{figure} 
\includegraphics[width=0.5\textwidth]{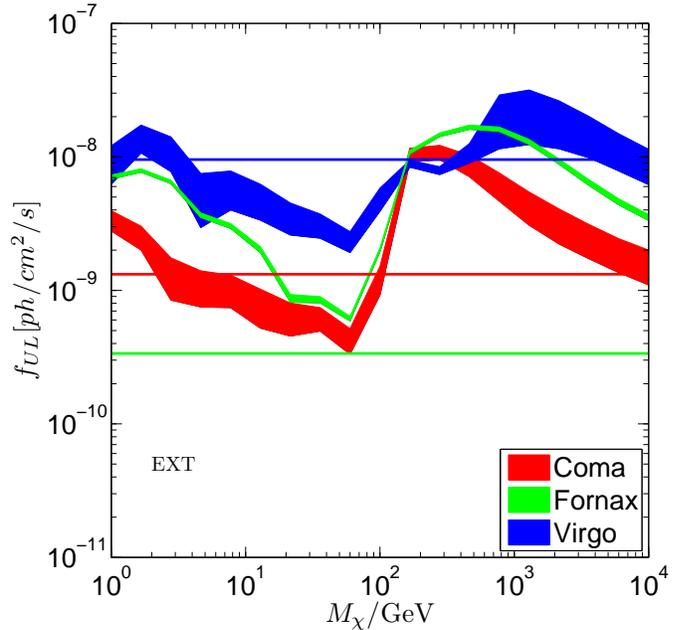} 
\caption{DM annihilation flux upper limits in the \mumu channel for the EXT model. 
 Line styles are as in Fig.~\ref{fig_Fluxbb}.} 
\label{fig_Fluxmm} 
\end{figure}

\begin{figure} 
\includegraphics[width=0.5\textwidth]{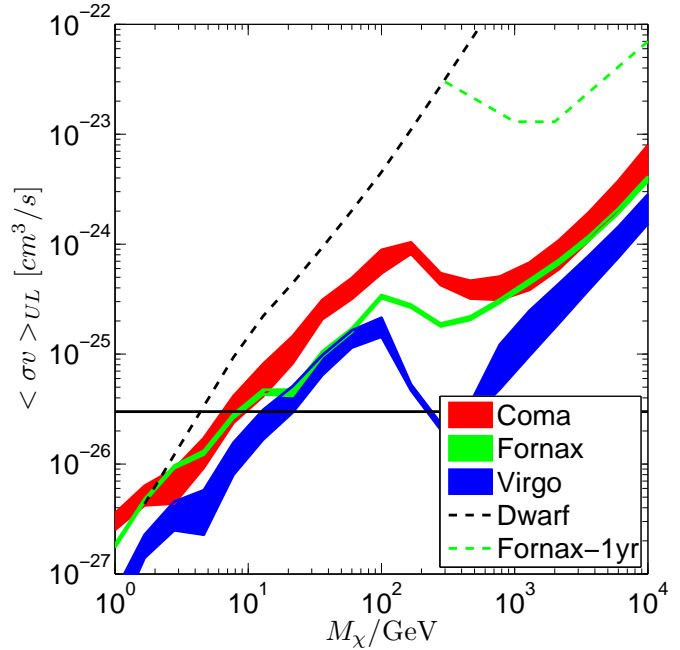} 
\caption{Upper limits for the DM annihilation cross-section in the 
\mumu channel. Line styles are as in Fig.~\ref{fig_SigVbb}, 
but only the EXT results are shown. The green dashed line is 
the 11-month Fermi result \citep{FermiCluster} for Fornax 
while the black dashed line is the dwarf galaxy constraint \citep{Dwarf}, 
both for the \mumu channel.} 
\label{fig_SigVmm} 
\end{figure}

As have been seen in section~\ref{sec_bb}, 
the EXT model places tighter constraints on the cross-section than the 
PT model, and 
is the fiducial model expected from recent simulations. Therefore 
from now on we will only show results for the EXT model. 
The flux and cross-section upper limits for DM annihilating 
through the \mumu channel are plotted in 
Fig.~s~\ref{fig_Fluxmm}  and~\ref{fig_SigVmm}. The predicted flux upper 
limits for Coma and Virgo are still comparable to the CR level, with 
Fornax having much lower CR emission. The inferred cross-section falls 
below the canonical value for DM particle masses less than 10~GeV. 
Note the discontinuity in the upper limits 
around 100~GeV which reflect the 
transition from the prompt annihilation dominated regime to the IC 
emission dominated regime in the photon spectrum.

\subsubsection{The \tautau channel}\label{sec_tautau} 
\begin{figure} 
\includegraphics[width=0.5\textwidth]{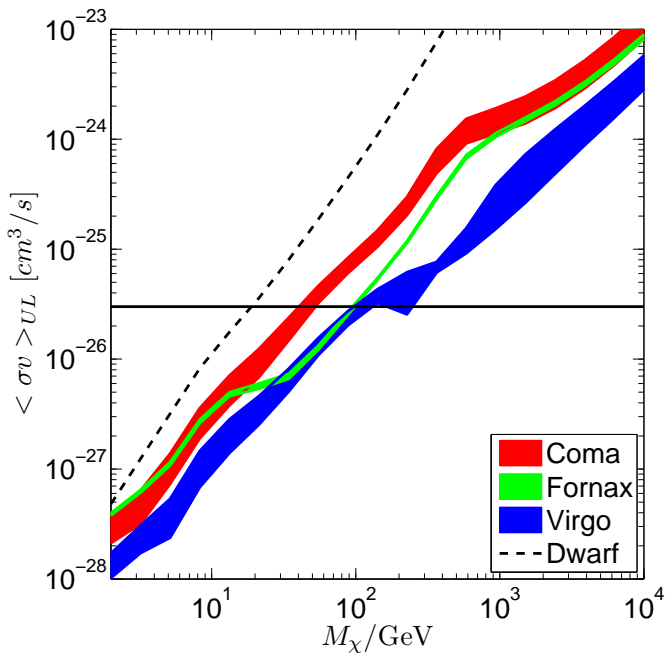} 
\caption{Upper limits for the DM annihilation cross-section in the 
  \tautau channel. Line styles are as in Fig.~\ref{fig_SigVbb}, but 
  only the EXT results are shown. The black dashed line is the dwarf 
  galaxy constraint. 
\citep{Dwarf}} 
\label{fig_SigVtt} 
\end{figure}

In Fig.~\ref{fig_SigVtt} we show the cross-section upper 
limits for the \tautau channel. This is the primary component of the 
leptonic model used by \cite{GalCen} to fit the excess gamma-ray 
emission in the Galactic centre region. 

\subsection{Allowing for an undetected point source population}\label{sec_correct}

Although we have detected five new point sources in the 45-month data in the 
region of our three clusters, it is still necessary to account for 
population of still undetected point sources. When no unknown point 
sources are present, the probability of measuring a certain value of 
$TS$ when the null hypothesis is true is given by the probability that 
Poisson fluctuations in the photon counts for the null model exceed 
some value.  When a population of undetected point sources is present, 
the Poisson fluctuations become correlated and it is easier for the 
same amplitude of fluctuations to result in a given value of $TS$. In 
this case, the distribution of $TS$ no longer follows the $\chi^2$ 
distribution, as predicted by Wilk's theorem, because the data no longer 
follow a pure Poisson distribution from which the likelihood function 
is constructed.

Allowing for the presence of undetected point sources in the data will 
lead to weaker upper limits. We obtain these by performing 
Monte-Carlo simulations to re-calibrate the significance corresponding 
to a given value of $TS$. In the simulations we include the GAL and 
2FGL sources, but we split the EG component into two parts: a 
population of undetected point sources and a residual smooth EG 
component, such that the sum of the two is consistent with the 
standard EG component. We consider a benchmark model for the 
undetected point source population which is close to the model derived 
by \citet{PTpop} and which contributes 14\% of the EG background. Details of 
the simulations may be found in Appendix~\ref{sec_CPD}.  A standard 
likelihood analysis is then performed on the simulated data in order 
to derive appropriate values of $TS$ for an assumed DM or CR 
component.

With the introduction of the undetected point source population, the 
distribution function for $TS$ is found to be roughly described by 
$\chi^2(TS/b)/2$. That is, the significance of a given value of $TS$ 
is approximately reduced by a factor of $b$ compared to the 
significance of the same value of $TS$ in the absence of the 
undetected point source population. For a DM component, we find $b\sim 
2$ for Coma and $b \sim 3$ for Virgo and Fornax. The $b$ factor is not 
sensentive to the adopted DM spectrum. For the CR models, we find 
$b\sim 2$ for Coma and Fornax, and $b\sim 3$ for Virgo.

In order to obtain new limits from the corrected TS, let us first 
consider the likelihood function that has been maximized over 
all the nuisance parameters. Expanding around the maximum likelihood 
value of the parameter $N$ to leading order, we have 
\begin{equation} 
\ln L(N_0)=\ln L(N)-\frac{1}{2} H (N_0-N)^2, 
\end{equation} 
where $H=-\frac{d^2 \ln L}{d N^2}=\frac{1}{\sigma_N^2}$ is the Hessian maxtrix. 
The 95\% upper limit is calculated from $\ln L(UL)-\ln L(N)=-1.35$, so that 
\begin{equation} 
UL=N+1.64\sigma_N. 
\end{equation} 
Note that $\frac{N}{\sigma_N}=\sqrt{TS}$. Similar equations hold 
for the improved background model, with $UL$, $\sigma_N$ and $TS$  replaced 
by $UL'$, $\sigma_N'$ and $TS'$ respectively, assuming that there is no bias in the best-fit parameters. 
The 95\% upper limit is then corrected for the undetected point source 
fluctuations according to 
\begin{equation} 
UL'=\frac{\sqrt{TS}+1.64\sqrt{TS'/TS}}{\sqrt{TS}+1.64}UL, 
\end{equation} 
where $UL$ and $TS$ are the upper limit and likelihood ratio from the 
standard analysis, while $UL'$ and $TS'$ are the corrected upper limit 
and likelihood ratio. For $b=3$, the increase in the upper limit is at 
most 70\%.

\begin{figure*} 
\includegraphics[width=0.33\textwidth]{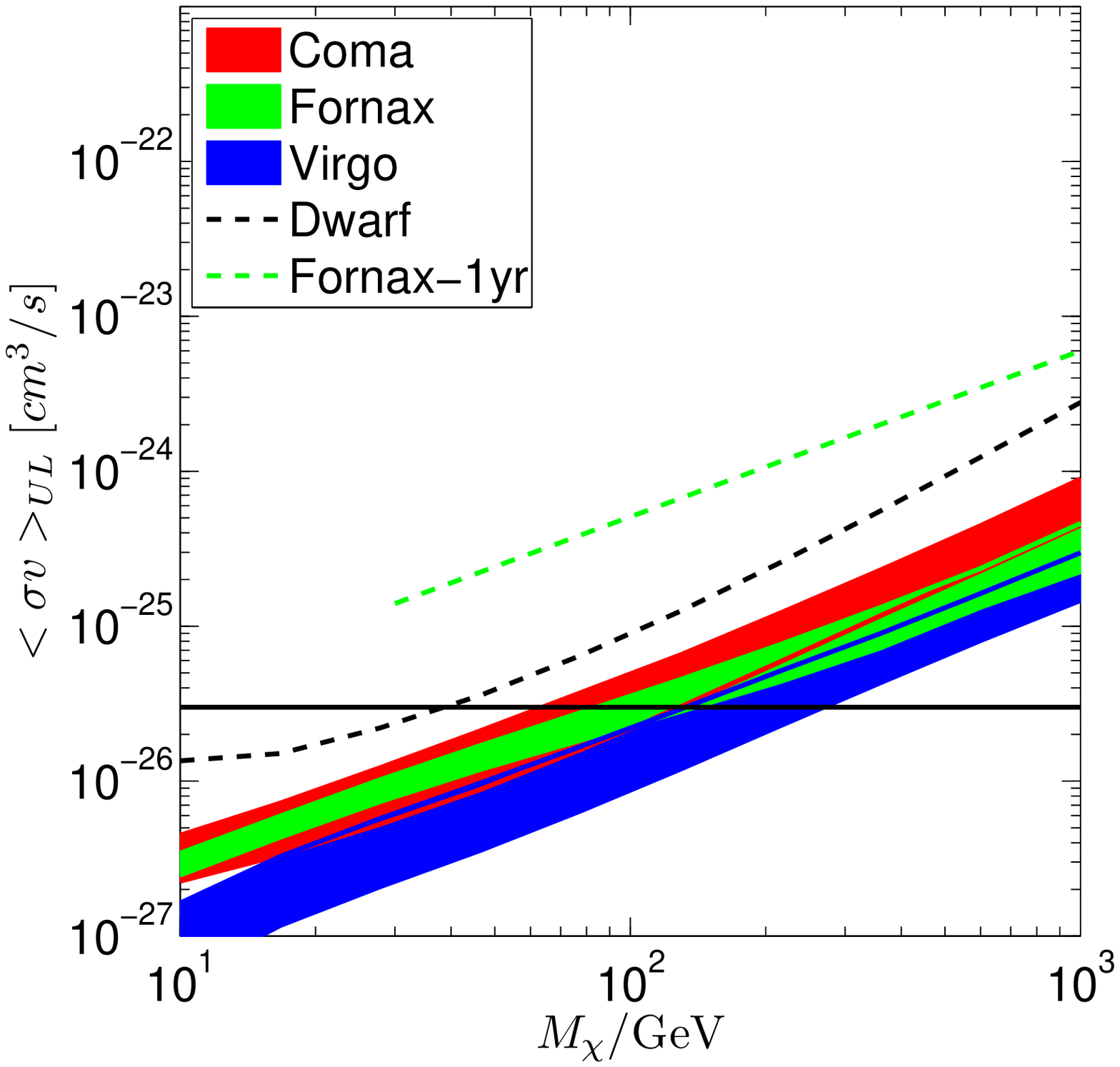}%
\includegraphics[width=0.33\textwidth]{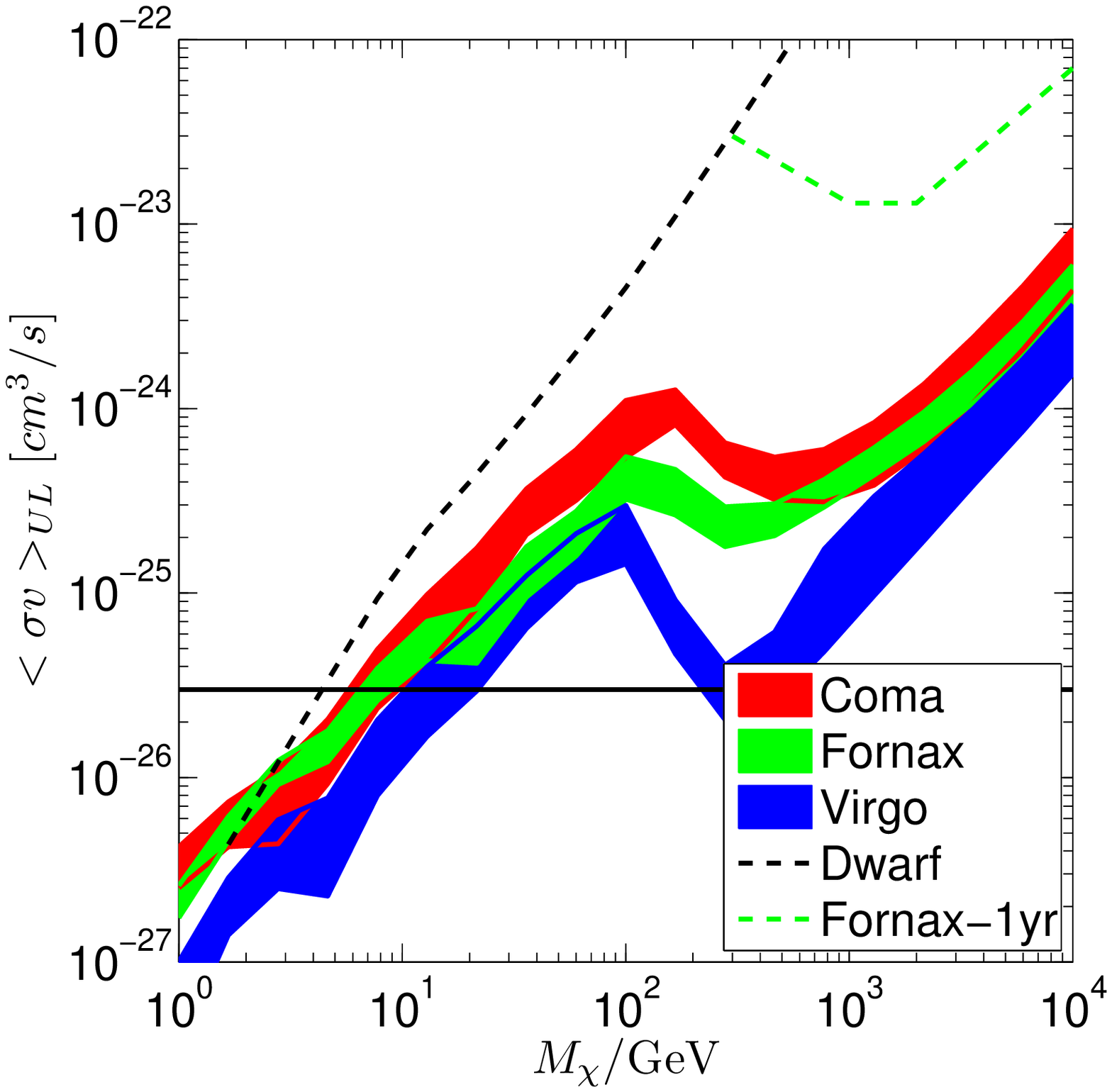}%
\includegraphics[width=0.33\textwidth]{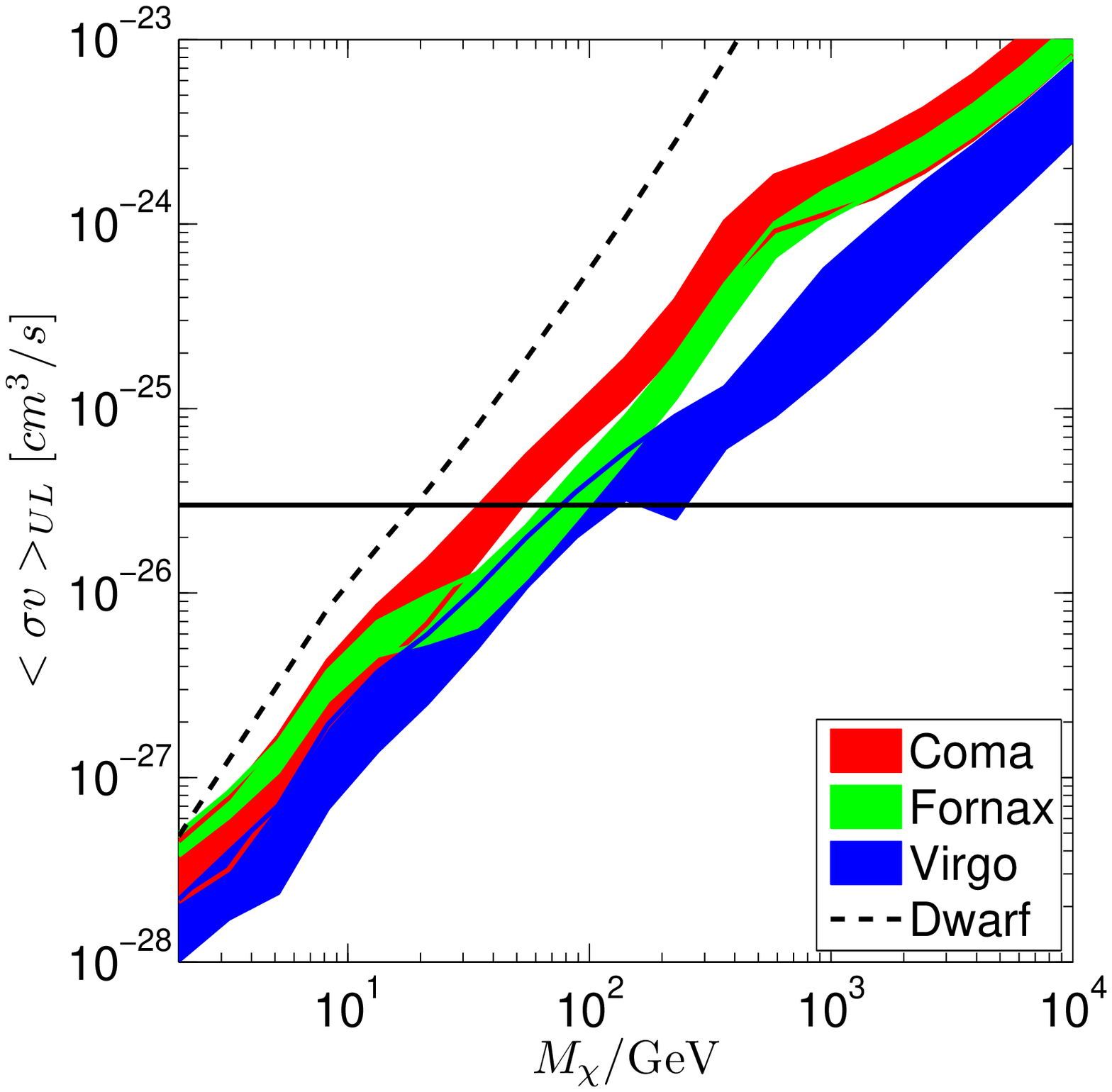} 
\caption{Upper limits for the DM annihilation cross-section in the 
\bbbar (left), \mumu (middle), and \tautau (right) channels, after including the effect of 
undetected point sources. Line styles are as in Fig.~\ref{fig_SigVbb}, 
but only the EXT results are shown. Note that the lower bounds of each band are still 
determined by the results without including undetected point sources 
in the analysis.}\label{fig_SigVCPD} 
\end{figure*}

In Fig.~\ref{fig_SigVCPD}, we show the corrected dark matter 
annihilation cross-section upper limits adopting $b=2$ for Coma and 
$b=3$ for Virgo and Fornax. The corrected TS and upper limits for CR 
models are listed in the last two columns of Table~\ref{table_CR}.

\section{Discussion and conclusions} \label{sec_discussion}

We have performed maximum likelihood fits to the 3-year Fermi-LAT data 
for three galaxy clusters: Coma, Fornax and Virgo. We fit models 
which, in addition to point sources and galactic and extragalactic 
backgrounds, include emission due to dark matter (DM) annihilation and 
cosmic rays (CR). For the former, we assume both a point source and 
the theoretically predicted extended distribution of gamma rays in 
three generic annihilation channels, the \bbbarnoc, \mumu and \tautau 
channels. When searching for a dark matter signal, we experiment with 
different treatments of the CR component. In the traditional Fermi 
analysis, the extragalactic background (EG) is assumed to be a smooth 
component. In this work we have also investigated a more realistic EG 
model where a fraction of the EG emission comes from a population of 
undetected point sources.

Performing a standard likelihood analysis we obtain the following results:

\begin{enumerate} 
\item In all three clusters and for the four different treatments of 
  CR we have implemented, no significant detection of DM emission is 
  obtained. We set upper limits on the flux and cross-section of DM 
  annihilation in the three clusters we have investigated. 
  Uncertainties in the CR component have only a mild effect on the 
  upper limits: for the different CR models, the DM upper limit 
  constraints agree to within a factor of two.

  Models in which the DM annihilation emission has the extended 
  profile predicted by cosmological simulations \citep{Gao} have 
  higher flux upper limits than models in which this emission is 
  assumed to be a point source. Due to the large luminosity 
  enhancement, of order of 1000, by emission from subhalos, the upper 
  limits on the annihilation cross-section for extended models are at 
  least 100 times lower than those for point source models. Our 
  cross-section constraints are much tighter than those from an 
  analysis of clusters using the 11-month data \citep{FermiCluster}, 
  mostly because we take into account the effect of subhalos. Our 
  constraints are also tighter than those from a joint analysis of 
  Milky Way dwarf galaxies \citep{Dwarf,DwarfFermi}.

  Our new limits exclude the thermal cross-section for $M_\chi\lesssim 
  100$~GeV for \bbbar and \tautau final states, and for 
  $M_\chi\lesssim 10$~GeV for \mumu final states. We note that the 
  annihilation cross-section in dark matter halos need not be the 
  standard thermal cross-section of supersymetric models. In cases 
  where the cross-section is velocity dependent, for example, through 
  p-wave contributions at freeze-out (see e.g., \citet{SUSYDM}), one 
  can easily have a different average cross-section. {We emphasize that
  there is still a large uncertainty our adopted annihilation profile,
  which depends on a significant extrapolation of the resolved subhalo population
  by more than 10 orders of magnitude in mass. Taking this into account, the thermal 
  cross-section, however, could still be reconciled with the data by 
  assuming a larger cutoff mass in the WIMP power spectrum, thus 
  reducing the contribution from subhalos and hence the $J$ 
  factor. Since the total enhancement from subhalo emission scales as 
  $b\propto M_{cut}^{-0.226}$ \citep{Volker}, a cut-off mass of 
  $10^{-4}\msun$, rather than our assumed $10^{-6}\msun$, would be 
  sufficient to increase the cross-section limits by a factor of~3.}

\item Assuming no DM annihilation radiation, the gamma ray data for 
  Coma and Virgo already set significant constraints on the CR 
  level. For Virgo, the data are consistent with the predictions of 
  the analytic CR model proposed by \citet{Pinzke1} and \citet{Pinzke2} 
  while, for Coma, the data place an upper limit that is a factor of 
  two below the analytical prediction, indicating either an 
  uncertainty in model parameters such as halo mass, gas density and 
  maximum shock injection efficiency, $\zeta_{p,max}$, or a 
  peculiarity of the CR emission in Coma. If attributed to 
  $\zeta_{p,max}$, the upper limit on the normalization parameter, 
  $\alpha_{CR}$, translates into an upper limit on $\zeta_{p,max}$ of 
  0.3, assuming a linear form for $g(\zeta_{p,max})$. This is 
  consistent with the estimates obtained independently by 
  \citet{Zimmer11} for Coma using Fermi data and by the 
  \citet{Perseus} for the Persus cluster using MAGIC observations. If 
  interpreted as an error in the halo mass, a reduction in mass by a 
  factor of $1.6$ is required to reconcile the model with the upper 
  limits, assuming a simple CR luminosity scaling relation, $L_\gamma 
  \propto M_{200}^{1.46}$ \citep{Pinzke1}, or a factor of $4.3$ 
  according to Eqn.~\ref{eq_CRR} in the case when the gas density 
  profile is fixed from X-ray observations. For Fornax, the 
  zero-significance of a CR component is consistent with the low level 
  predicted by the model.

\item Five new point sources with $TS>25$ in Virgo and Fornax have 
  been detected in the 45-month data. Ignoring these new point sources 
  results in a $\sim 5\sigma$ detection for a DM component in Virgo, in 
  contrast to a $\sim 3\sigma$ detection when account is taken of 
  these point sources. 
\end{enumerate}

In addition to the standard likelihood analysis, we have also investigated 
a model in which the EG component includes a population of undetected 
point sources whose number-flux relation extrapolates smoothly that 
of the detected sources. Using Monte-Carlo simulations, we 
find that the standard Fermi likelihood analysis could 
overestimate the TS of extended emission by a factor of $2-3$, and 
underestimate the upper limits by up to 70 percent. Adopting this 
more realistic EG model yields slightly looser upper limits, but does 
not quantitatively change any of the above conclusions.  Still, it 
should be kept in mind that these corrections are derived from 
simulations assuming a particular distribution for the point source 
population. It is too computationally expensive to explore the parameter space of point 
source populations with Monte-Carlo simulations. A more detailed and more 
general analytical study of the effect of undetected point sources 
will be presented elsewhere (Han et. al., in preparation).

In our analysis we have allowed the parameters of 2FGL point sources 
lying within the cluster virial radius to vary. This accounts for 
possible corrections to the 2FGL parameters in the presence of a DM or 
a CR component, while also avoiding the risk of refitting sources 
lying near the boundary of the data region with less accuracy. The 
parameters of highly variable sources are also kept free since the 
2FGL parameters for these sources would be the average during a 2 year 
period whereas here we have 45 months of data. However, we also tried 
keeping all the point sources fixed or allowing the parameters of all 
the point sources within the data region to vary during the 
fitting. We find that this freedom in the treatment of the point 
sources has little impact on the DM model fits.

The cluster annihilation luminosity scales roughly linearly with halo 
mass, with the shape of the profile being almost independent of halo 
mass or concentration when expressed in terms of the normalized radius 
$r/R_{200}$. We investigate the effect of mass uncertainties in 
Appendix~\ref{sec_mass}. We have also checked that the different 
energy cuts assumed in our analysis and in that of \citet{Huang} have 
no effect on the derived upper limits. We are able to reproduce the 
upper limits on the annihilation cross-section of \citet{Huang} for 
the test case of the Fornax cluster with 3-year data, after adopting 
the same instrument response function and correcting for slightly 
different assumed subhalo contributions.

The CR model used in this analysis is still subject to 
improvement. This model is derived from simulations which, 
unavoidably, make simplifying assumptions. For example, the 
simulations only consider advective transport of CR by turbulent gas 
motions but there are other processes such as CR diffusion and 
streaming which may flatten the CR profiles \citep{Ensslin11}. In 
particular, if the CR diffusion is momentum dependent this will 
entangle the spectral and spatial profile of CR and modify the 
morphology as well as the spectrum of the CR emission, thus 
invalidating our basic assumption that $\alpha_{CR}$ is the only free 
parameter. There could also be CR injected from AGN which are not 
accounted for in the current model.

Although we have not detected DM annihilation emission in our small 
cluster sample, the signal-to-noise ratio can potentially be enhanced 
by stacking many clusters. Such an analysis was recently carried out 
by \citet{Huang}, but the signal-to-noise was degraded because of 
their assumption of an NFW annihilation profile. These authors 
considered an extended subhalo-dominated annihilation profile but only 
for individual clusters, not for the stack. Their stacked analysis 
placed looser constraints on DM annihilation emission than their 
analysis of individual clusters, presumably because the use of an 
inappropriate theoretical profile resulted in the different clusters 
yielding inconsistent results. Thus, it is clearly worth repeating the 
joint analysis with the ``correct'' subhalo-dominated profile. It is 
also tempting to extend the search for DM annihilation using 
multi-wavelength data, from the radio to very high energy gamma-rays {and even
in the neutrino channel\citep{Neutrino}}, where different systematics are expected for different bands.

\section*{Acknowledgments} 
We thank Shaun Cole, Jie Liu, Yu Gao, John Lucey, Anders Pinzke, 
Christoff Pfrommer, Dan Hooper, Neal Weiner, Douglas Finkbeiner, 
Gregory Dobbler, Louie Strigari, Christoph Weniger, Savvas 
Koushiappas, and Fabio Zandanel for helpful discussions. JXH 
acknowledges the support on software issues from Tesla Jeltema and the 
Fermi science support team, especially Elizabeth C. Ferrara, Jeremy 
S. Perkins, Dave Davis and Robin Corbet. JXH is supported by the 
European Commissions Framework Programme 7, through the Marie Curie 
Initial Training Network Cosmo-Comp (PITNGA-2009-238356), and 
partially supported by NSFC (10878001, 11033006, 11121062) and by the 
CAS/SAFEA International Partnership Program for Creative Research 
Teams (KJCX2-YW-T23). CSF acknowledges a Royal Society Wolfson 
research merit award and an ERC Advanced Investigator grant. The 
calculations for this work were performed on the ICC Cosmology 
Machine, which is part of the DiRAC Facility jointly funded by STFC, 
the Large Facilities Capital Fund of BIS, and Durham University. This 
work was supported in part by an STFC rolling grant to the ICC.  
The work of D.M. is supported in part from the SCOPES project IZ73Z0 
128040 of Swiss National 
Science Foundation, grant No CM-203-2012 for young scientists of 
National Academy of Sciences 
of Ukraine, Cosmomicrophysics programme of the National Academy of 
Sciences of Ukraine and by the State Programme of Implementation of Grid Technology in Ukraine.

\bibliographystyle{\mybibstyle} \bibliography{Fermi_ref}
\numberwithin{figure}{section}
\appendix
\section{Detection of New Point Sources}\label{sec_NewPT}

We model the new point sources assuming power-law spectra. For a given
pixel, we calculate the $TS$ value for an assumed new point source
centered on that pixel. The $TS$ calculation is performed using the binned
method in the \texttt{pyLikelihood} tool, with a null model which
includes the GAL and EG components and all the 2FGL sources within 15
deg of each cluster, but with the parameters of the 2FGL sources
fixed. Around each cluster, we carry out a first scan of all the
pixels within the cluster virial radius (and within 4 deg around
Coma) using a pixel size of $0.2$ deg.

Regions with a peak $TS>16$ are identified as potential locations of
new point sources. We then scan each potential point source region
using 10 times smaller pixels. The calculated $TS$ map is then
interpolated with cubic splines down to 0.002 deg$/$pixel. The value
and location of the $TS$ peak is taken as the $TS$ and position for a new
point source, if the peak $TS>25$. In case several peaks are
clustered, we first extract the primary $TS$ peak, then scan for lower
$TS$ peaks by including the newly detected sources into the null
model. In our sample, no secondary peaks survive this iterative
examination to be identified as new point sources.

\begin{figure*}
\includegraphics[width=0.5\textwidth]{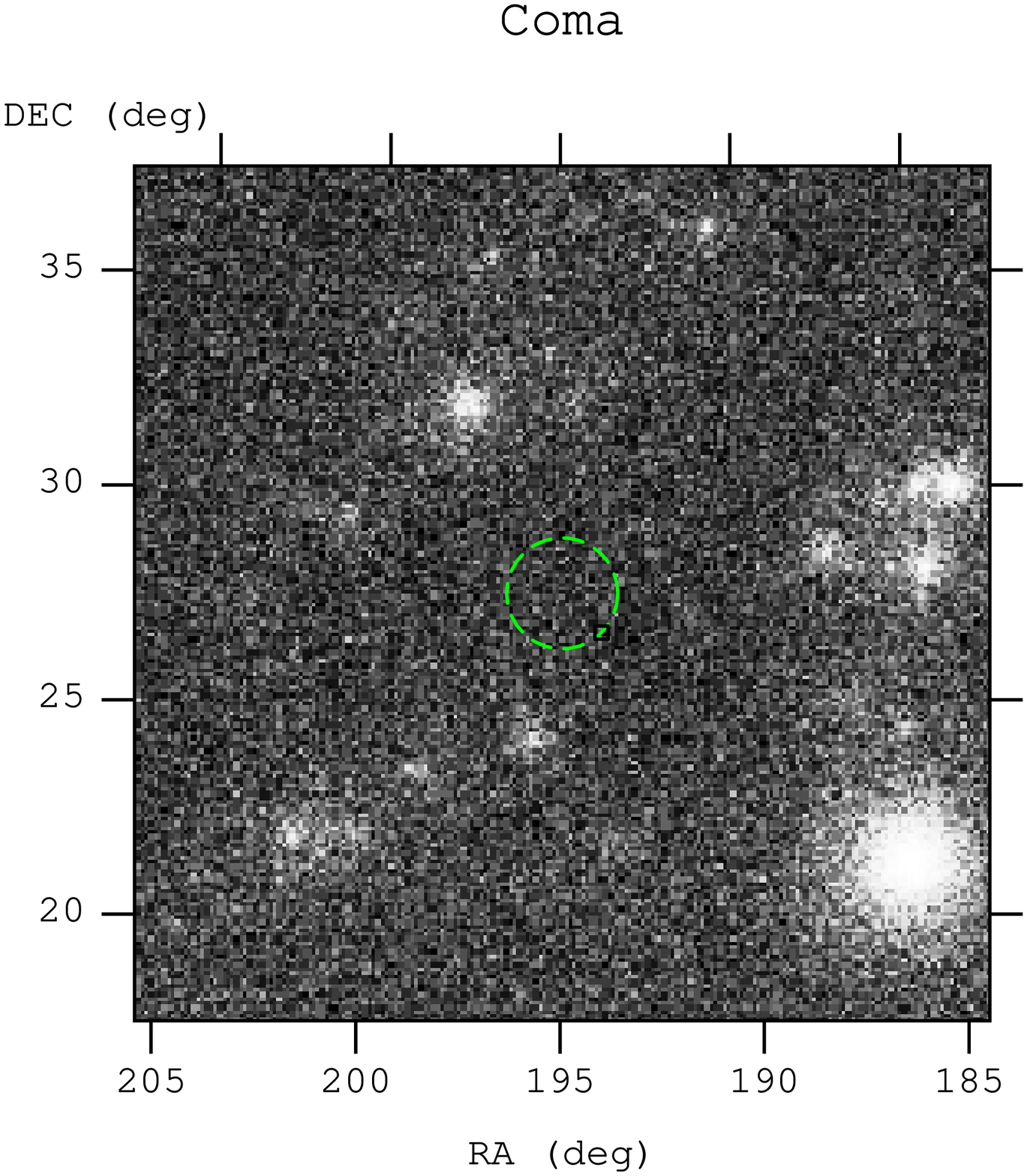}%
\includegraphics[width=0.5\textwidth]{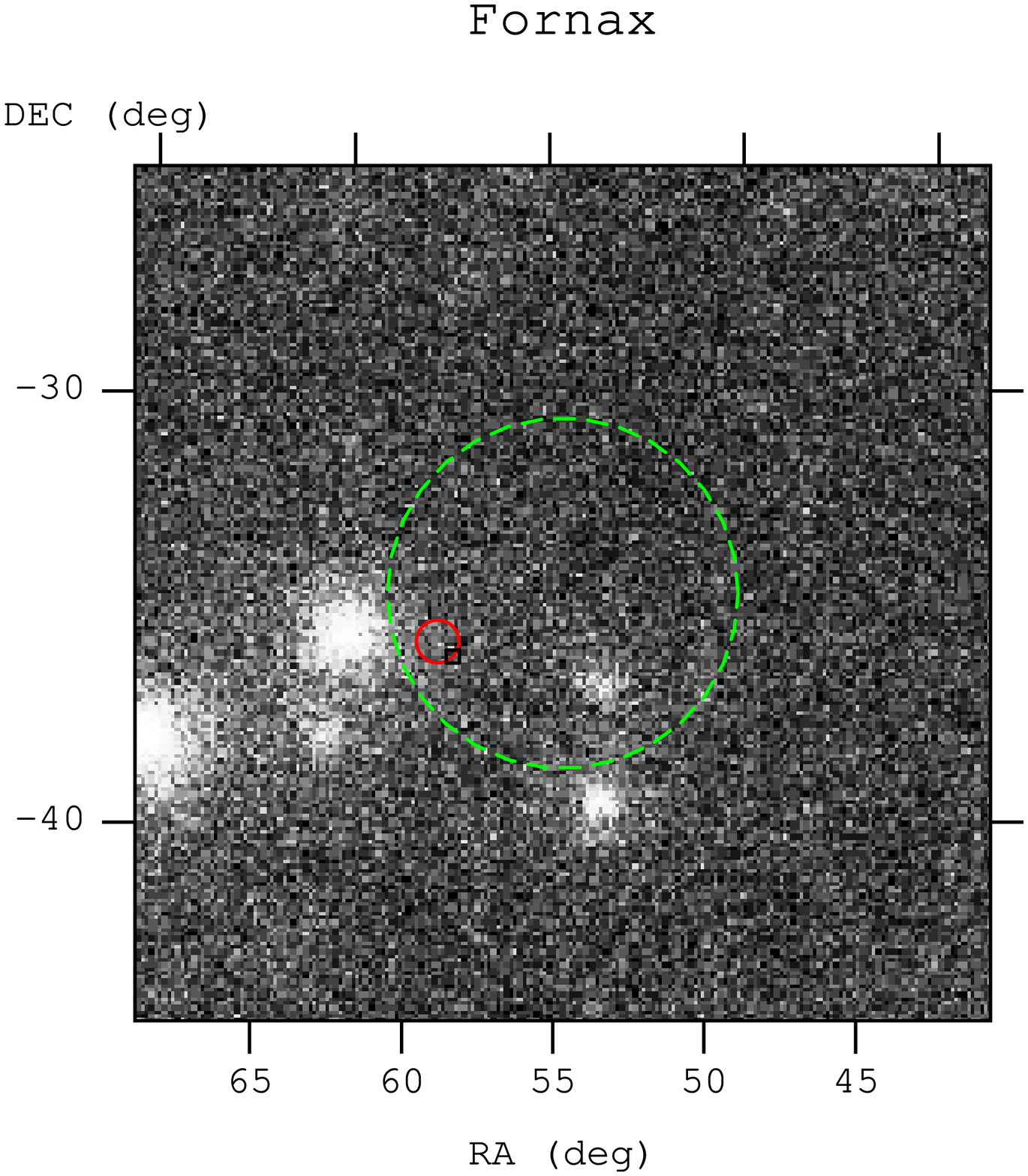}
\caption{Integrated gamma-ray images in the Coma (left) and Fornax
(right) cluster regions. The green dashed circle marks the virial
radius of the cluster. Each image covers $20\times20 \ \rm{deg}^2$ with
a pixel size of 0.1 deg, constructed from the 3-year Fermi-LAT data
applying the data cuts described in the main text. The small solid circle in Fornax marks the position of a newly detected point source.}
\label{fig_CMaps}
\end{figure*}

The new point sources are listed in Table~\ref{table_NewPT}, and ploted in Figure~\ref{f_components} and \ref{fig_CMaps}. Sources in Virgo and Fornax are prefixed by ``V" and ``F" in their names respectively. 
 None of these new sources show significant variability when binned over monthly scale. The last column of Table~\ref{table_NewPT} shows possible associations of astrophysical sources with these new detections, which are found to lie within the 2$\sigma$ confidence region of the detections.
\begin{table*}
\caption{Newly detected point sources}\label{table_NewPT}
\begin{tabular}{|c|c|c|c|c|c|c|c|}
\hline Name & TS & RA (deg) & DEC (deg) & Flux~($10^{-9}$ ph $\cdot$ cm$^{-2}$ s$^{-1}$) & Spectral Index\tablenotemark{a} & Seperation (deg)\tablenotemark{b} & Possible Association \\
\hline  V1 & 32.5 & 190.920 & 16.194 & $5.9\pm 1.4$  & $-2.3\pm 0.2$ &  4.96 & LBQS 1241+1624 \\
\hline  V2 & 31.8 & 185.698 & 11.116 & $3.7\pm 1.0$  & $-2.0\pm 0.2$ &  2.31 & [VV2006] J122307.2+110038  \\
\hline  V3 & 31.6 & 184.066 & 9.456  & $2.3\pm 0.8$  & $-1.9\pm 0.2$ & 4.58 &  2MASX J12160619+0929096 \\
\hline  V4 & 30.5 & 185.894 & 8.286 & $1.6\pm 0.7$  & $-1.8\pm 0.2$ & 4.42  & SDSS J122321.38+081435.2\\
\hline  F1 & 26.3 & 58.300 & -36.386 & $0.9\pm 0.6$ & $-1.7\pm 0.3$ &  3.17 & [VV98b] J035305.1-362308\\
\hline
\end{tabular}\\
\tablenotetext{a}{Photon spectral index $\beta$ for $dN/dE\propto E^{\beta}$.}
\tablenotetext{b}{Distance to cluster centre.}
\end{table*}

\section{Monte-Carlo simulation of undetected point source populations}\label{sec_CPD}

To model the undetected population we adopt the following model based
on the results of \citet{PTpop}. Each point source is assumed to have
a power-law spectrum defined by two parameters: flux and spectral
index. The spectral index distribution is modeled as a Gaussian of
mean $\mu=2.36$ and $\sigma=0.27$. The flux and spectral index
are assumed to be independent. The differential
number density of undetected point sources is assumed to be given by
\begin{equation}
\frac{dN}{dSd\Omega}=A(\frac{S}{S_b})^{-\beta}.
\end{equation}
We adopt $S_b=6.6\times 10^{-8} \ {\rm ph \ cm}^{-2} \ {\rm s}^{-1}$,
$\beta=1.58$ and $A=4.1\times 10^8 \ {\rm cm^2\ s\ Sr^{-1}}$, as
derived from Table~4 of \citet{PTpop}. Since the total number of point sources
diverges for $\beta>1$, we cut off the flux distribution at $S_{\rm
  min}=1\pow{-11} \ {\rm ph\ cm^{-2}\ s^{-1}}$. Due to the dependence
of the detection efficiency on flux and spectral shape, there is no
obvious cutoff in the maximum flux of undetected sources.  We take
$S_{\rm max}=1\pow{-8}\ {\rm ph \ cm^{-2} \ s^{-1}}$ as the detection
threshold which corresponds to a detection completeness of $\sim
50\%$, comparing 2FGL source counts and the model. This implies an undetected point source flux of 14\% of the
standard EG background, consistent with the results of
\citet{PTpop}. The synthetic spectrum of these undetected point
sources is then subtracted from the standard EG template to yield a
residual EG template for the simulation.


%
%

We perform 750 independent realizations of the 15~deg Virgo region in
the presence of undected point sources. For each realization, we
generate mock data in the following steps:
\begin{itemize}
\item Generate a Poisson random number for the total number of
  undetected point sources within 15~ deg.
\item For each point source, generate a random spectral index and a
  random flux according to the distributions specified above. Also,
  generate random coordinates for the point source according to a
  uniform distribution on the sky.
\item Feed these point sources and the 2FGL point sources within 15
  deg, as well as the GAL and remaining EG components, to
  \texttt{gtobssim}.
\end{itemize}

\begin{figure*}
\includegraphics[width=0.5\textwidth]{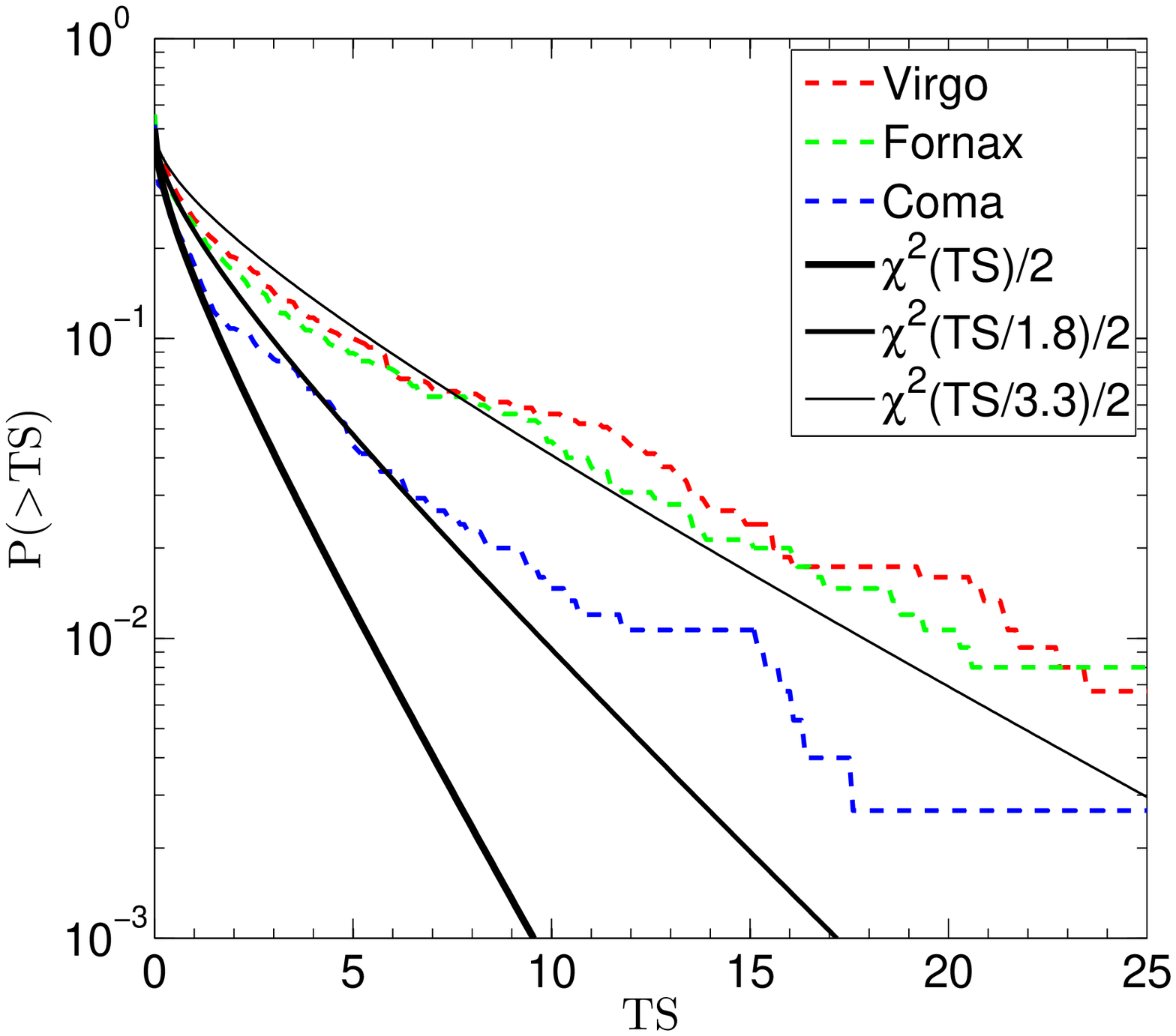}%
\includegraphics[width=0.5\textwidth]{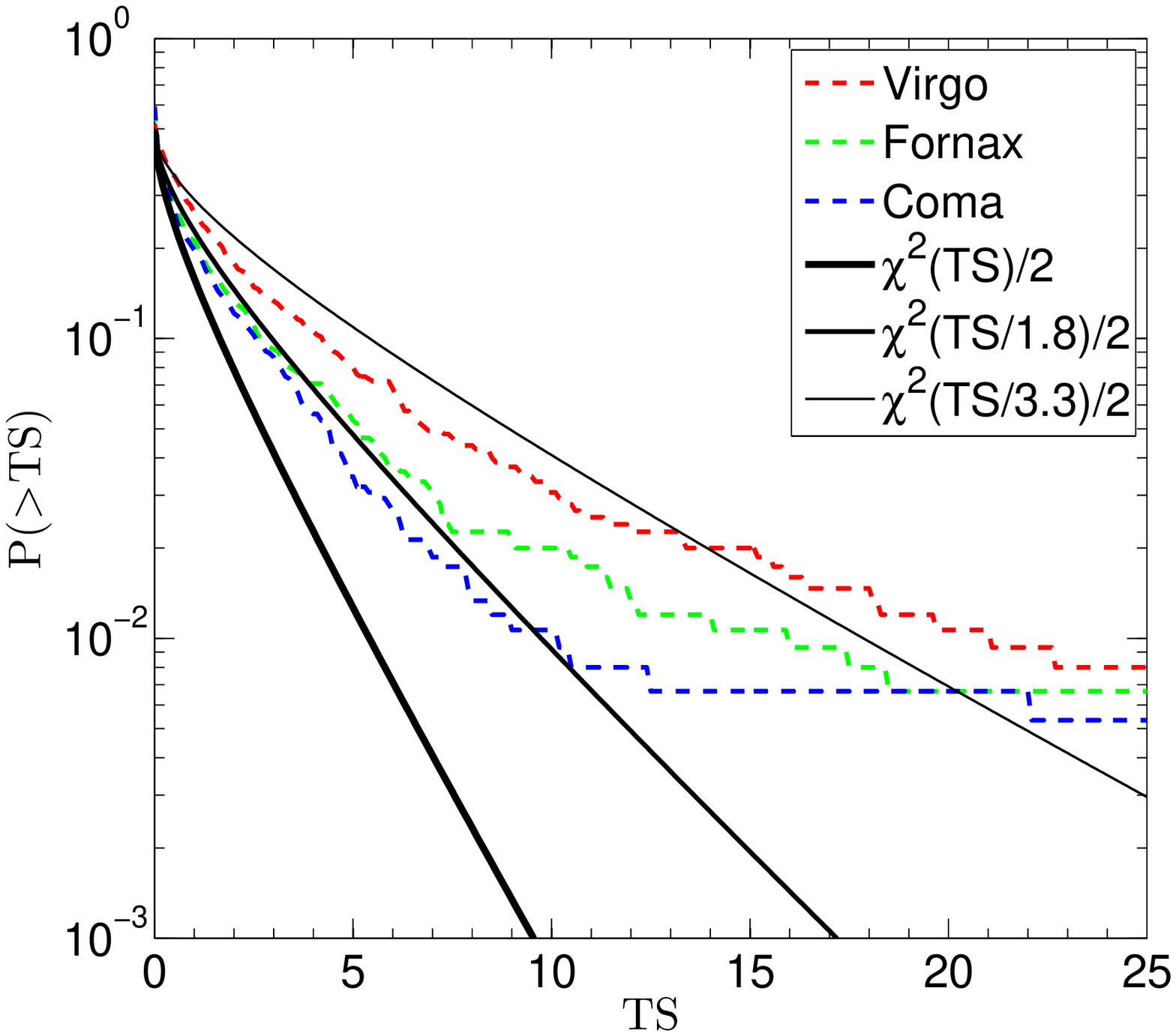}
\caption{Distribution of $TS$ from simulated data which include a
  population of undetected point sources. Left: the distribution of
  $TS$ for DM-only models, where the DM particle mass is taken to be
  $\sim 30$~GeV and the DM follows the EXT cluster profile.  Right:
  the distribution of $TS$ for CR-only models. In each panel the
  dashed lines show the distribution extracted from the simulations
  for three cluster models and the solid lines show a rescaled version
  of the standard cumulative $\chi^2$ distribution.}\label{fig_TSCPD}
\end{figure*}

The standard likelihood analysis is then applied to the simulated
data without including any of the randomly generated point sources in
the model. Here we only consider the CR-only and the DM-only
models. In Fig.~\ref{fig_TSCPD} we show the cumulative probability
distribution of $TS$ values. Simple scaled versions of the standard
$\chi{^2}(TS)/2$ distributions can roughly describe the $TS$ distribution
and provide the simplest way to convert the fitted $TS$ to the standard
$\chi^2$-distributed $TS$.

\section{Gamma-ray images for Coma and Fornax}

In this Appendix we show gamma-ray images for the Coma and Fornax
cluster regions. The corresponding image for Virgo is shown in
Fig.~\ref{f_components}.

\section{Semi-Analytic formula for the Cosmic Ray induced gamma-ray emission}
Here we summarize the relevant equations for calculating the CR
induced gamma-ray emission in galaxy clusters as derived by
\citet{Pinzke1} and \citet{Pinzke2}. The CR induced photon source
function from pion decay can be decomposed as:
\begin{equation*}
\frac{dN_{\gamma}}{dtdVdE}=A(r)s(E).
\end{equation*}
The spatial part is given by:
\begin{equation}\label{eq_CRR}
A(r)=((C_{200}-C_{centre})(1+(\frac{r}{R_{trans}})^{-\beta})^{-1}+C_{centre}){\rho_{gas}(r)^2},
\end{equation}
with
\begin{eqnarray}
C_{centre}&=5\times 10^{-7}\\
C_{200} & = 1.7 \times 10^{-7} \times (M_{200}/10^{15}M_\odot)^{0.51}\\
R_{trans} & = 0.021 R_{200}\times (M_{200}/10^{15}M_\odot)^{0.39}\\
\beta & =1.04 \times (M_{200}/10^{15}M_\odot)^{0.15}
\end{eqnarray}
The spectrum is given as:
\begin{eqnarray}
s(E) &
=g(\zeta_{p,max})D_{\gamma}(E_{\gamma},E_{\gamma,break})\frac{16}{3m_p^3c}\nonumber\\
     &\times \sum_{i=1}^3 \frac{\sigma_{pp,i}}{\alpha_i}
(\frac{m_p}{2m_{\pi^0}})^{\alpha_{i}}\Delta_i[(\frac{2E_\gamma}{m_{\pi^0}c^2})^{\delta_i}+(\frac{2E_\gamma}{m_{\pi^0}c^2})^{-\delta_i}]^{-\frac{\alpha_i}{\delta_i}},
\end{eqnarray}
with $\Delta= (0.767,0.143,0.0975)$, $\alpha=(2.55,2.3,2.15)$,
$\delta_i\simeq 0.14\alpha_i^-1.6+0.44$. Here $m_p$ is the proton
mass, $m_{\pi^0}$ the neutral pion mass and $c$ the speed of
light. The maximum shock acceleration efficiency is chosen to be
$\zeta_{p,max}=0.5$ so that $g(\zeta_{p,max})=1$. The term
$D_\gamma(E_\gamma,E_{\gamma,break})$ describes the diffusive CR
losses due to escaping protons as
\begin{equation}
 D_\gamma(E_\gamma,E_{\gamma,break})=[1+(\frac{E_\gamma}{E_{\gamma,break}})^3]^{-1/9}.
\end{equation}
The proton cut-off energy is
\begin{equation}
E_{p,break}\approx \frac{10^8}{8} {\rm GeV} (\frac{R_{200}}{1.5Mpc})^6.
\end{equation}
The energy $E_{p,break}$ is related to the photon cut-off energy,
$E_{\gamma,break}$, through the momentum relation $P_{\gamma}\approx
\frac{P_p}{8}$. The effective cross-section for proton-proton
interactions is given by:
\begin{equation}
\sigma_{pp,i}\simeq 32(0.96+e^{4.42-2.4\alpha_i}) \mathrm{mbarn}.
\end{equation}
The gas density is fitted with multiple beta-profiles:
\begin{equation}
\rho_{gas}=\frac{m_p}{X_H X_e}\lbrace\sum_i
n_i^2(0)[1+(\frac{r}{r_{c,i}})^2]^{-3\beta_i}\rbrace^{1/2},
\end{equation}
where $X_H=0.76$ is the primordial hydrogen mass fraction and
$X_e=1.157$ is the ratio of electron and hydrogen number densities in
the fully ionized intracluster medium, with parameter values for
$n_i(0)$, $r_{c,i}$ and $\beta_i$ listed in TABLE VI of
\citet{Pinzke2}.

\section{Effect of mass uncertainties in Virgo}\label{sec_mass}
\begin{figure}
\includegraphics[width=0.5\textwidth]{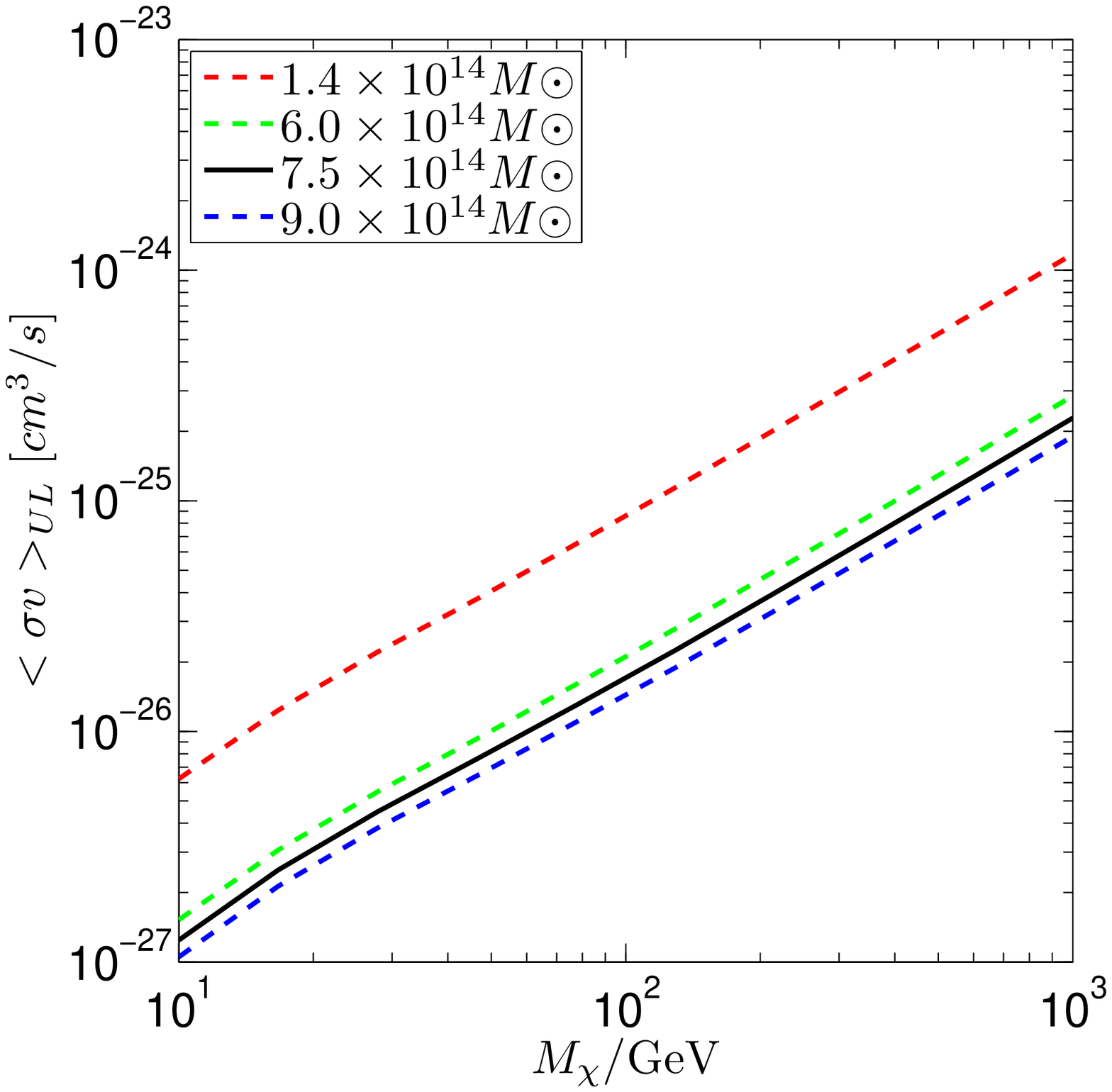}
\caption{Upper limits on the cross-section for DM annihilating into the \bbbar
  channel in the no-CR model.  Different lines correspond to different
  adopted values for the mass of the dark matter halo of the Virgo
  cluster, as labelled. No allowance for undetected point sources has
  been made in this figure.}
\label{fig_MassErr}
\end{figure}

We adopt a virial mass for Virgo of $7.5\pm1.5\times10^{14}M\odot$,
as estimated by \cite{TS84} from an analysis of the infall pattern of
galaxies around Virgo. This value is consistent with other dynamical
measurements
\citep{Smith,Hoffman,Tonry2000,VirgoMass,VirgoInfall}. Mass
estimates from X-ray gas modelling tend to give somewhat lower values
\citep{Bohringer04,Schindler,Urban}. Thus, in addition to our adopted
mass uncertainty from \citet{TS84}, as an extreme case, we consider
also a value of $1.4\times10^{14}M\odot$, obtained by scaling the
X-ray estimate to the virial radius \citep{Urban}. In
Fig.~\ref{fig_MassErr} we show the effect of adopting these
different masses on the upper limits for DM
annihilation in the \bbbar channel. Since the flux upper limit is
insensitive to slight changes in the profile shape and thus in the mass,
while the luminosity (or integrated $J$ factor) scales linearly with
mass, the cross-section upper limits are expected to be roughly
inversely  propotional to mass. This is indeed the case in
Fig.~\ref{fig_MassErr}, where $<\sigma v>_{UL} \propto
M_{200}^{-0.9}$.
